 \documentstyle[prd,aps,eqsecnum,floats,epsfig,12pt]{revtex}
\draft
\begin{document}
\date{17 December 1996}
%
%
\title{Exclusive diffractive processes 
 and the quark substructure of mesons}
%
\author{M.A.~Pichowsky\footnotemark[1]\footnotemark[2] and 
	T.-S.H.~Lee\footnotemark[1]}
\address{
   \footnotemark[1] 
	Physics Division, Argonne National Laboratory,
	Argonne, IL 60439-4843, USA 
		\vspace*{2mm}\\
   \footnotemark[2]
	Department of Physics \& Astronomy, University of Pittsburgh, 
	Pittsburgh, PA 15260, USA
}
\date{\today}
\maketitle
\begin{abstract}
Exclusive diffractive processes on the nucleon are investigated within a 
model in which the quark-nucleon interaction is mediated by Pomeron exchange
and the quark substructure of mesons is described within a framework based on
the Dyson-Schwinger equations of QCD. 
The model quark-nucleon interaction has four parameters which are completely
determined by high-energy $\pi N$ and $K N$ elastic scattering data.
The model is then used to predict vector-meson electroproduction observables.  
The obtained $\rho$- and $\phi$-meson electroproduction cross sections 
are in excellent agreement with experimental data. 
The predicted $q^2$ dependence of $J/\psi$-meson electroproduction
also agrees with experimental data.
It is shown that confined-quark dynamics play a central role in determining
the behavior of the diffractive, vector-meson electroproduction cross
section.  In particular, the onset of the asymptotic $1/q^4$ behavior of the
cross section is determined by a momentum scale that is set by the
current-quark masses of the quark and antiquark inside the vector meson.
This is the origin of the striking differences between the $q^2$ dependence
of $\rho$-, $\phi$- and $J/\psi$-meson electroproduction cross sections
observed in recent experiments.
\end{abstract}
%
%
\renewcommand{\-}{\!-\!}
\renewcommand{\+}{\!+\!}
\newcommand{\g}{\gamma}  
\newcommand{\sfrac}[2]{\mbox{\footnotesize $\frac{#1}{#2}$}} 
\newcommand{\intd}[1]{\int\!\!\!\frac{{\rm d}^4\!#1}{(2\pi)^4} }
\newcommand{\be}{\begin{equation}\ }
\newcommand{\ee}{\end{equation}}
\newcommand{\ba}{\begin{eqnarray}\ }
\newcommand{\ea}{\end{eqnarray}}
\newcommand{\nn}{\nonumber\ }        
\newcommand{\Fig}[1]{Fig.~\protect\ref{#1}}
\newcommand{\Figs}[1]{Figs.~\protect\ref{#1}}
\newcommand{\Sec}[1]{Sec.~\protect\ref{#1}}
\newcommand{\Ref}[1]{Ref.~\protect\cite{#1}}
\newcommand{\Tab}[1]{Table~\protect\ref{#1}}
\newcommand{\Eq}[1]{Eq.~(\protect\ref{#1})}
\newcommand{\Eqs}[1]{Eqs.~(\protect\ref{#1})}
\newcommand{\Refs}[1]{Refs.~\protect\cite{#1}}
\newcommand{\scdot}{\!\cdot\!}
%
%
\sloppy
\section{Introduction}
\label{Sec:Intro}

At high energies, the differential cross sections for hadron-hadron
elastic scattering are forward peaked and fall off exponentially in
the  small-momentum-transfer region.  
The magnitudes of these cross sections increase very slowly with energy 
at a rate that seems to be independent of the hadrons involved. 
The mechanism responsible for this behavior was originally identified 
as {\em Pomeron exchange} within Regge phenomenology
\cite{Collins,Irving}. Since the advent of Quantum Chromodynamics
(QCD) as the underlying theory of strong interactions, there has been
considerable effort towards the 
development of a description of Pomeron exchange directly in terms of
QCD. However, a satisfactory description has not yet been obtained.  
The objective of this work is to develop an understanding of the role 
played by the quark substructure of hadrons in diffractive processes
as a step towards providing a description of Pomeron exchange in terms
of the quark and gluon degrees of freedom of QCD\@.
In particular, we explore the consequences of nonperturbative-quark
dynamics, such as confinement and dynamical chiral symmetry breaking,  
in determining observables that are accessible in current experiments
at the Deutsches Elektronen-Synchrotron (DESY), 
Thomas Jefferson National Accelerator Facility (TJNAF) and 
Fermi National Accelerator Laboratory (FNAL).
The availability of high quality data from these facilities provides
a means to explore the dynamical content of Pomeron exchange which, in
turn, has motivated recent theoretical effort in this direction.

It is generally believed that the underlying mechanism
responsible for Pomeron exchange is multiple-gluon exchange. 
This idea was first investigated by Low within the Bag model
\cite{Low} and by Nussinov in \Ref{Nussinov}.
The simplest multiple-gluon exchange requires at least two gluons,
since all hadrons are color singlets.  
As there are many different ways in which two gluons can be
exchanged between two hadrons, a straightforward calculation of
two-gluon exchange is difficult.
For example, in pion-nucleon elastic scattering, there are several
ways in which the quarks inside the two hadrons can propagate and
exchange gluons. One contribution arises when the quark inside the pion
exchanges two gluons with a quark inside the nucleon. Another
possibility is that  the two exchanged gluons interact with two
different quarks inside the nucleon.  
There are other possibilities as well.
To further complicate matters, the exchanged gluons can interact with
each other. 
Such a calculation cannot at present be carried out without making
additional approximations.   

A simplification was introduced by Donnachie and Landshoff
\cite{DL84}, who proposed that the Pomeron couples to the nucleon like
an isoscalar photon.   
This led them to consider a model in which multiple-gluon exchange is 
replaced by a Regge-like Pomeron exchange whose coupling to the
nucleon is described in terms of a nucleon isoscalar electromagnetic
(EM) form factor.   
In our work, we employ a similar approach, thereby avoiding
the complexities associated with the quark substructure of the nucleon
so that we may focus on exploring the dynamics of Pomeron exchange and
its relation to the quark substructure of mesons.  
The model of \Ref{DL84} has also been applied to 
$\rho$-meson electroproduction \cite{DL87} with the simplifying
assumption that the quark propagators inside the $\rho$ meson can be  
factorized out of the quark-loop integration,
replaced by constituent-quark propagators 
and evaluated at a single momentum.   
A ramification of employing such a procedure is the loss of some of
the momentum dependence originating from the quark substructure of the
vector meson. 
Consequently, agreement with experimental data requires the
introduction of a quark-Pomeron form factor which is, in fact,
dependent on the type of vector meson produced (e.g. $\rho$, $\phi$,
$J/\psi$) \cite{Laget}.   
However, when the quark substructure of the vector meson is properly 
accounted for, {\em no} quark-Pomeron form factor is required to
obtain agreement with $\rho$-meson electroproduction data
\cite{Pichowsky}. The extent to which the same is true for other vector
mesons is addressed in this work. 

The first objective of this investigation is to construct a model
which restores some of the important features of QCD and
nonperturbative-quark propagation to the study of Pomeron exchange and
the role of the quark substructure of mesons in diffractive processes.
Some nonperturbative aspects of QCD, like quark and gluon confinement, are
expected to play important roles in exclusive processes \cite{Pennington}. 
We find that incorporating such features into the model considered
herein, naturally leads to predictions in agreement with the observed $q^2$
dependence of $\rho$-, $\phi$- and $J/\psi$-meson 
electroproduction cross sections, within a single framework and {\em
without} introducing quark-Pomeron form factors. 

The second objective of this investigation is to use the constructed
model to determine the extent to which Pomeron exchange can be
interpreted within QCD as multiple-gluon exchange.
There have been some experimental observations that support this
notion and numerous theoretical investigations on the subject,
starting with those in Refs.~\cite{Low,Nussinov}.
One such experimental observation is the near flavor 
{\em independence} of high-energy meson-nucleon elastic scattering
amplitudes.  
If Pomeron exchange is predominantly gluons, then flavor dependence
observed in the diffractive processes of light quarks, should arise
mostly from the quark substructure of the hadrons involved. 
A satisfactory resolution of this outstanding problem requires 
that cross sections for diffractive processes are calculated
from a model in which the perturbative and nonperturbative
properties of quarks, confined within hadrons, are accounted for.
That is, the model employed should have the properties expected
from nonperturbative QCD and provide a good description of the
low-energy properties of hadrons, such as their EM form factors and
magnetic moments.   
Therefore, in this work, we use the confined-quark propagator and
Bethe-Salpeter (BS) amplitudes obtained from phenomenological studies
of hadrons based on the Dyson-Schwinger equations (DSEs) of QCD. 

The DSEs are an infinite set of coupled integral equations that relate
all of the $n$-point functions of a quantum field theory to each other.  
The generating functional of a quantum field theory being completely
determined once all of the $n$-point functions are known.  Hence,
knowledge of the $n$-point functions completely specifies the dynamics
of the theory. 
The simplest of these is the 2-point, quark-propagator DSE,
which describes how dressing of the quark propagator is dynamically
generated by the quark's interactions with its own gluon field.
To obtain a solution of this integral equation requires knowledge of
other $n$-point functions, which in turn, satisfy their own DSEs.  
In general, the kernel of a DSE that determines an $n$-point function
contains at least one $m$-point function with $m > n$.
This illustrates the self-coupling between the equations that
necessarily entails a truncation scheme in order to obtain a finite
and tractable set of equations.
Quantitative studies of such truncated systems of DSEs have had
considerable success; a review of this body of work may be found in
\Ref{RobertsDSE}.    

The quark propagator employed herein is based on that obtained in a
numerical study of the DSEs using a model-gluon propagator
\cite{FrankGluon}. 
The obtained quark propagator exhibits dynamical chiral symmetry
breaking and confinement, both of which are essential to provide a
good description of nonperturbative QCD phenomena.  
The significance of their role in diffractive processes is discussed
later.  
Here, we only mention that the confined-quark propagator and BS
amplitudes obtained from phenomenological studies of the DSEs provide
an excellent description of processes involving mesons. 
Some of these processes include
the $\pi$- and $K$-meson EM form factors \cite{Burden}, 
$\pi\pi$ scattering \cite{RobertsPi}, 
$\rho\omega$ mixing \cite{Tandy}, 
the $\gamma \pi^* \rightarrow \pi \pi$ form factor \cite{Alkofer},
the $\gamma^*\pi\gamma$ transition form factor \cite{FrankGPiG} and 
various EM and weak decays of the $\pi$ and $K$ mesons 
\cite{Burden,RobertsPi,Kalinovsky}.
The use of a confined-quark propagator distinguishes our approach from 
previous studies of Pomeron exchange and we find that it is essential
in obtaining a consistent description of vector-meson
electroproduction for both large and small values of the photon
momentum squared, $q^2$.  

With the confined-quark propagator and Bethe-Salpeter amplitudes taken
from phenomenological DSE studies, the parameters of the quark-nucleon
Pomeron-exchange interaction, introduced herein, are {\em completely
determined} by $\pi N$ and $K N$ elastic scattering data. 
The model interaction is then used to investigate vector-meson
electroproduction, which is 
currently the focus of great experimental and theoretical interest.
We first consider exclusive $\rho$-meson electroproduction and then
$\phi$- and $J/\psi$-meson electroproduction.

The $q^2$ dependence of the vector-meson electroproduction cross
sections obtained is particularly interesting. 
For large $q^2$,  
these cross sections obey an asymptotic power law, proportional to $1/q^4$.
The transition to this asymptotic behavior is determined by the
current-quark masses of the quark and antiquark inside the vector meson.  
It follows that the electroproduction cross sections of vector mesons 
comprised of heavy quarks exhibit this transition at a higher $q^2$
than those comprised of light quarks. 
This feature is demonstrated by an application of our model to
$\rho$-, $\phi$- and $J/\psi$-meson electroproduction.  
Our predictions for the $q^2$ dependence of $\rho$-, $\phi$- and
$J/\psi$-meson electroproduction cross sections are in excellent
agreement with the recent experimental data.

  At present several theoretical investigations of diffractive vector-meson
electroproduction appear in the literature. Some of these employ a
perturbative methods in QCD 
\cite{Brodsky,Ryskin,Ginzburg,Dosch,Martin,Frankfurt,CollinsFrankfurt} 
while others (including the present study) employ nonperturbative methods
\cite{DL84,DL87,Laget,CudellRoyen,Jenkovszky}. 
The range of applicability of perturbation theory in determining
diffractive vector-meson electroproduction amplitudes at HERA energies is
still under discussion.  
In \Ref{Brodsky}, it is argued that only the electroproduction of {\em
longitudinally} polarized vector mesons should be considered and even
then only at large photon momentum squared, $q^2$.
It has also been suggested that perturbative methods would only be reliable
for the production of heavy-quark vector mesons \cite{Ginzburg}.  
This is supported by the experimental data for $J/\psi$-meson
electroproduction which appear to be well described using such methods
\cite{Ryskin},  
while nonperturbative effects, like the relative, ``Fermi-momentum''
between the quark and antiquark inside the vector meson, must be included
when considering light-quark vector meson electroproduction \cite{Frankfurt}.

Although there are similarities between these perturbative approaches and
the approach considered herein, there is an important difference.
For large photon momentum squared, the perturbative description of
diffractive electroproduction proceeds by a mechanism in which the 
photon fluctuates into a quark-antiquark pair that then scatters
elastically from the proton target and ultimately forms the on-mass-shell
vector meson.   
It is assumed that the quark and antiquark each carry a large fraction of
the total photon momentum and can therefore be treated as freely
propagating, on-mass-shell partons in order to determine the 
$\bar{q}q$-proton elastic scattering amplitude.
In contrast to this, we find that the presence of the vector meson bound
state amplitude constrains the
momentum flow in the quark loop so that one of the quark propagators that
couples to the quark-photon vertex is hard and the other is soft.
(This feature was first observed in \Ref{DL87} and is shown in 
\Sec{Sec:Vector} to be true in our model as well.)  
An important ramification of this is that the asymptotic limit of the
quark-photon vertex, $\gamma_{\mu}$, does not provide the dominant
contribution to the electroproduction cross section for any value of $q^2$.
The Ward-Takahashi identity requires that {\em both} fermion legs must
carry large momentum before the perturbative limit is reached;
the presence of a vector-meson Bethe-Salpeter amplitude keeps this from
happening in diffractive vector-meson electroproduction.

As an example of the importance of the nonperturbative contributions to the
quark-photon vertex, we 
make predictions for the ratio of the longitudinal to
transverse vector-meson electroproduction cross sections, $R =
\sigma_L / \sigma_T$. 
We obtain a value of $R$ that is much smaller than other models and
smaller than the values extracted from $\rho$-meson electroproduction
polarization data under the assumption of $s$-channel helicity
conservation. 
Replacing the dressed quark-photon vertex by a bare vertex, $\gamma_{\mu}$,
reduces the value of $R$ which gives an indication of the sensitivity of
this observable to nonperturbative contributions in our model.

The organization of this paper is as follows.
In \Sec{Sec:Model}, we give a precise formulation of the quark-nucleon 
Pomeron-exchange interaction to be considered in this work.
The parameters of the model are determined in \Sec{Sec:Elastic}
by considering $\pi N$ and $K N$ elastic scattering data. 
The model is applied to vector meson electroproduction 
and the results are presented in \Sec{Sec:Vector}. 
Finally, in \Sec{Sec:Conclusion}, we present our conclusions.

\section{A Pomeron-exchange model of the quark-nucleon interaction}
\label{Sec:Model}

We begin by constructing a phenomenological model of the interaction 
between a quark, confined within a hadron, and an on-mass-shell
nucleon.  
We only consider processes in which {\em no} flavor is exchanged
between the quark and nucleon, so that in the high-energy and
small-momentum-transfer region, 
this interaction can be parameterized in terms of {\em Pomeron
exchange}. 
The most general form of the quark-nucleon Pomeron-exchange
interaction may be represented as the shaded box in \Fig{Fig:qqNN} and
written in momentum space as 
\ba
\lefteqn{{\cal I}(k,k^{\prime};p_1,p_2) = } 
\label{GenPomDef} \\
& & \sum_{\mu \nu f}
\bar{q}^f(k^{\prime}) \Gamma_{\mu}^f q^f(k) \, 
{\bf G}_{\mu \nu}(k,k^{\prime};p_1,p_2)\,  
\bar{u}_{m^\prime}(p_2) \Gamma_{\nu} u_{m}(p_1),
\nn\
\ea
where: $u_{m}(p_1)$ and $\bar{u}_{m^\prime}(p_2)$ are the Dirac spinors 
for the incoming and outgoing nucleons; 
$q^f(k)$ and $\bar{q}^f(k^{\prime})$ are elements of the Grassmann
algebra for a quark of flavor $f = u, d, s,...$; 
$\Gamma_{\nu}$ is a matrix in the Dirac space of the nucleon; and 
$\Gamma_{\mu}^f$ is a matrix in the Dirac and flavor spaces of the
quark.  
The momentum dependence of the exchange mechanism and the
quark-Pomeron and nucleon-Pomeron couplings is given by  
the amplitude ${\bf G}_{\mu \nu}(k,k^{\prime};p_1,p_2)$.   
In \Eq{GenPomDef} and throughout this work, we have suppressed color
and Dirac indices. We employ the Euclidean metric, 
$\delta_{\mu \nu} = {\rm diag}(1,1,1,1)$  
and hermitian Dirac matrices that satisfy: 
$\{\g_{\mu},\g_{\nu}\} = 2 \delta_{\mu \nu}$. 

The Pomeron-exchange model of the quark-nucleon interaction,
employed throughout this work, is constructed from \Eq{GenPomDef} by 
introducing the following three simplifying assumptions:

%
\begin{figure}[t]
\centering
{\ \epsfig{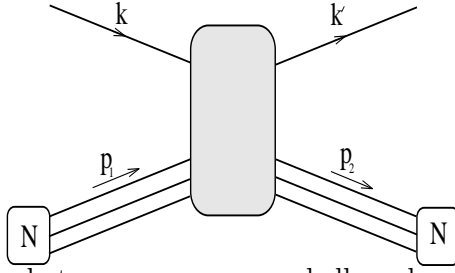}}
\caption{
General interaction between an on-mass-shell nucleon and confined
quark, given by \Eq{PomDef}.} 
\label{Fig:qqNN}
\end{figure}
%
%
$(i)\;$ Motivated by the observation that the differential cross
sections of $\pi N$ and $KN$ elastic scattering have similar $t$
dependence,  
we assume that the quark-Pomeron coupling, $\Gamma_{\mu}^f$, can be
factorized into two independent matrices,  
\be
 \Gamma_{\mu}^f \longrightarrow \Gamma_{\mu} \beta_{f},
\label{Assumption1}
\ee
where $\Gamma_{\mu}$ is a constant matrix in the Dirac space of the
quarks and $\beta_f$ is a constant.
This leads to the simplification that apart from a flavor-dependent
overall multiplicative constant $\beta_f$, the quark-Pomeron coupling
is flavor {\em independent}. 

The use of a flavor-dependent coupling constant, $\beta_f$,
is {\em not} inconsistent with the identification of
Pomeron exchange as a multiple-gluon exchange.  
In contrast to the bare quark-gluon vertex (which is flavor
independent) the nonperturbative dressing of the quark-gluon vertex
{\em required} by the Slavnov-Taylor identity, introduces a nontrivial
flavor dependence in $\Gamma_{\mu}^f$.

Therefore, \Eq{Assumption1} incorporates the assumption that all of
the flavor dependence of the quark-Pomeron-exchange vertex can be
absorbed into a {\em constant}, $\beta_f$.  We will see that this 
minimal flavor dependence is sufficient to describe processes
involving light quarks. 
However, in \Sec{Sec:VectorJPsi}, we show it is too restrictive to
describe processes involving heavy quarks, like $J/\psi$-meson
electroproduction, and must be relaxed in this case.   

$(ii)\;$ We assume that the Dirac structure of the
quark-Pomeron-exchange coupling in \Eq{Assumption1} and the
nucleon-Pomeron-exchange coupling in \Eq{GenPomDef} can be taken
as $\Gamma_{\mu} = \g_{\mu}$. 
Although this assumption is overly simplistic and perhaps, dubious, it 
has been employed in previous work \cite{DL84,DL87,Laget,Pichowsky} and by
maintaining this assumption, we can directly compare our results to
these studies. 
Furthermore, the above assumption leads to approximate $s$-channel
helicity conservation which is often associated with Pomeron exchange. 

The dubious aspect of employing $\gamma_{\mu}$ as the 
quark-Pomeron-exchange vertex lies in the fact that $\gamma_{\mu}$ is
odd under charge conjugation, while Pomeron-exchange should be even.
In this work, the charge-conjugation parity of the
quark-Pomeron-exchange coupling is implemented by hand.  
In our approach, we can investigate the possibility of
a more sophisticated quark-Pomeron-exchange vertex, which has the
supposed symmetries of Pomeron exchange.  
Such a study is left to future work.  

$(iii)$ The quark-nucleon Pomeron-exchange interaction, 
${\bf G}_{\mu \nu}(k,k^{\prime};p_1,p_2)$,  
is assumed to be independent of the square of the four momenta of the
quarks, $k^2$ and $k^{\prime 2}$, and diagonal in the Lorentz indices.
Hence, we write: 
${\bf G}_{\mu \nu}(k,k^{\prime};p_1,p_2) 
= \delta_{\mu \nu} \tilde{G}(s,t)$
where $s =-(k + p_1)^2$ and $t = - (p_1 - p_2)^2$.
To make contact with previous Pomeron-exchange models
\cite{DL84,DL87},  
we introduce the following parameterization:
\be 
\tilde{G}(s,t) \equiv 3 \beta_u G(s,t) F_1(t),
\label{GDef1}
\ee
where,
\be
F_1(t) \equiv \frac{4 M_N^2 - 2.8 t}{4 M_N^2 - t} 
\frac{1}{(1 - t/t_0)^2}
\label{F1Def} ,
\ee
is the nucleon isoscalar EM form factor, $M_N$ is the nucleon mass and 
$t_0 = $ 0.7~GeV$^2$.
The appearance of the EM form factor, $F_1(t)$, in \Eq{GDef1},
suggests a similarity between photon and Pomeron exchange.
The difference between these two exchanges is provided by the
function, $G(s,t)$, which is parameterized as
\be
G(s,t) = (\alpha_1 s)^{\alpha_0+ \alpha_1 t},
\label{GDef}
\ee
and is reminiscent of the Pomeron-exchange Regge trajectory discussed 
in Refs.~\cite{Collins,Irving}.
Of course, by employing a value of $\alpha_0 > 0$, we recover the
observed behavior that cross sections of processes which proceed
through Pomeron exchange increase with energy, $\sqrt{s}$.

With the above assumptions, \Eq{GenPomDef} is reduced to 
\ba
I(k,k^{\prime};p_1,p_2) &=& 
[\sum_{f} \bar{q}^f(k^{\prime}) \beta_f \g_{\mu} q^f(k)] 
\, [3 \beta_u G(s,t) F_1(t)] 
\nn\
\\
& & \times [\bar{u}_{m^{\prime}}(p_2) \g_{\mu} u_m(p_1)]
.
\label{PomDef}
\ea
Although our model interaction, defined by \Eq{PomDef}, is similar in
appearance to that of other authors \cite{DL84,Laget},
our description of the quark substructure of mesons differs
significantly and plays a pivotal role in the processes considered.

Another difference between our approach and those of previous authors 
is our assumption that the quark coupling to the Pomeron-exchange
interaction is independent of the quark four momenta, $k^2$ and
$k^{\prime 2}$.
Other authors have found it necessary to weaken this assumption  by 
introducing an additional dependence on the quark momenta in the form
of a ``quark-Pomeron form factor'' \cite{DL87,Laget}. 
The scale associated with this form factor turns out to be 
flavor {\em dependent} \cite{Laget}, thereby violating our first
assumption, as well. 
In our approach, we do not introduce this additional momentum
dependence.  
In fact, we show that the flavor dependence of our predictions is
entirely determined by the overall, multiplicative constant $\beta_f$
and the quark substructure of the mesons involved.  
Having parameterized the quark-nucleon Pomeron-exchange interaction in 
\Eqs{GDef} and (\ref{PomDef}), we now proceed to determine the
parameters. 

Throughout this work we assume $SU(2)_{\rm flavor}$ is a good
symmetry, so that $\beta_d = \beta_u$.
Hence, the four parameters, $\alpha_0$, $\alpha_1$, $\beta_u$ and
$\beta_s$ completely determine the quark-nucleon Pomeron-exchange
interaction for the light, $u$, $d$ and $s$ quarks according to
Eqs.~(\ref{GDef}) and (\ref{PomDef}).  
In \Sec{Sec:Elastic}, these parameters are determined by considering
$\pi N$ and $K N$ elastic scattering. 

\section{Elastic scattering of pseudoscalar mesons and nucleons}
\label{Sec:Elastic}

The first application of our model is to obtain cross sections for
$\pi N$ and $KN$ elastic scattering.
This requires a description of the $\pi$- and $K$-meson
quark-antiquark bound states.
Such a description is provided for by phenomenological studies of the
DSEs of QCD. 

In \Ref{Burden}, confined, $u$-, $d$- and $s$-quark propagators and
$\pi$- and $K$-meson Bethe-Salpeter amplitudes were obtained and shown
to provide an excellent description of $\pi$- and $K$-meson
observables, including their EM form factors. 
These elements, having been previously determined in \Ref{Burden}, 
allow us to use the existing $\pi N$ and $KN$ elastic scattering data
to completely determine the parameters, $\alpha_0$, $\alpha_1$ and
$\beta_f$ ($f=u,d,s$),  in the quark-nucleon Pomeron-exchange
interaction of \Eq{GDef} and (\ref{PomDef}). 

Of the parameters $\alpha_0, \alpha_1$ and $\beta_f$ ($f=u,d,s$),
$\alpha_0$ is an observable, directly related to the universal $s$
dependence of hadron-hadron total cross sections. 
Its value can, therefore, be taken from the analysis of experimental
data given in \Ref{Cudell}.    
In this study, it was found that the data for $pp$ and $\bar{p}p$
total cross sections are best reproduced by the value of 
$\alpha_0 = 0.096^{+0.012}_{-0.009}$.
For our purposes, $\alpha_0 = 0.10$ is sufficient.
Our task is then to determine $\alpha_1$ and $\beta_f$ from $\pi N$
and $K N$ elastic scattering data.
 
Assuming $SU(2)_{\rm flavor}$ symmetry for the $u$ and $d$ quarks, we
set $\beta_d = \beta_u$.   
The parameters, $\beta_u$ and $\alpha_1$, are determined by
considering the $\pi N$ elastic scattering differential cross section and 
$\beta_s$ is determined by the {\em magnitude} of the $KN$ 
elastic scattering cross section.    
The $t$ dependence of the $K N$ differential cross section is the 
first prediction of our model.  In our approach, the $t$ dependence of
the quark-nucleon Pomeron-exchange interaction, $G(s,t)$ in \Eq{GDef},
is the {\em same} for
all processes involving light quarks.   
Therefore, the predicted differences in the
differential cross sections for $\pi N$ and $K N$ elastic scattering
are {\em entirely} due to the quark substructure of the $\pi$ and $K$
mesons. 
To our knowledge, this is the first work to correlate the $\pi N$ and
$KN$ differential cross sections and (as we will see) properly account
for the observed differences between them.
 
\subsection{Pion-nucleon elastic scattering}
\label{Sec:ElasticPi}
%
%
%
\begin{figure}[t]
\centering
{\ \epsfig{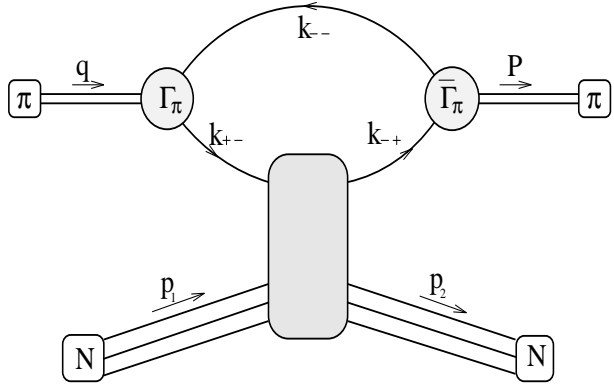}}
\caption{One of two diagrams that contributes to $\pi N$ elastic
scattering in impulse approximation.} 
\label{Fig:PiN}
\end{figure}
 
The $\pi N$ elastic scattering amplitude is described within our model
as Pomeron exchange between the nucleon and the quark or antiquark
inside the $\pi$ meson.  This is illustrated in \Fig{Fig:PiN}.  
Application of the quark-nucleon Pomeron-exchange interaction defined
by \Eq{PomDef} leads to the $\pi N$ scattering amplitude,
\ba
\lefteqn{\langle P ; p_2 m^{\prime} | T_{\pi N \rightarrow \pi N}| 
q ; p_1 m \rangle
= } \nn\ \\
& & 2 i \Lambda_{\mu}(q,P) \; [3 \beta_u F_1(t) G(s,t)] \; 
[\bar{u}_{m^\prime}(p_2) \g_{\mu} u_{m}(p_1)]
\label{TPiN}
.
\ea
Here, $\Lambda_{\mu}(q,P)$ describes the coupling of a $\pi$ meson to
the Pomeron-exchange interaction and the remaining terms were defined
in \Sec{Sec:Model}. 
In deriving the product form of \Eq{TPiN}, we have assumed that the
$s$ dependence of the  Pomeron-exchange interaction, $G(s,t)$, can be
evaluated from the external pion and nucleon momenta, so that $s = -(q
+ p_1)^2$ is the usual Mandelstam variable.
This approximation is commonly employed in Pomeron-exchange
models \cite{DL84,DL87,Laget}. It is adopted here to ensure that 
the $\Lambda_{\mu}(q,P)$ is a function of $q$ and $P$ only, which
reduces the number of integrations required to evaluate
$\Lambda_{\mu}(q,P)$.  
We have found that this approximation can, at most, lead to a 10\%
change to the normalization of the amplitude and has no observable effect
on the $s$ and $t$ dependence.  

For $\pi^+ N$ elastic scattering, $\Lambda_{\mu}(q,P)$ is comprised of
two terms, 
\be
\Lambda_{\mu}(q,P) = \Lambda^{u}_{\mu}(q,P) +
\Lambda^{\bar{d}}_{\mu}(q,P)
,
\ee
where
\ba
\Lambda^{u}_{\mu}(q,P) &=& N_c  
\,{\rm tr}\!\intd{k} 
S_u(k_{+-}) \Gamma_{\pi}(k\-\sfrac{1}{2}P)
\nn\ \\
& & \times S_d(k_{--}) \bar{\Gamma}_{\pi}(k\-\sfrac{1}{2}q) 
   S_u(k_{-+}) \beta_u \g_{\mu}
, \label{LambdaPiu} \\
\Lambda^{\bar{d}}_{\mu}(q,P) &=& N_c 
\,{\rm tr}\!\intd{k} 
S_d(k_{+-}) \bar{\Gamma}_{\pi}(k\-\sfrac{1}{2}P)
\nn\ \\
& & \times
S_u(k_{--}) {\Gamma}_{\pi}(k\-\sfrac{1}{2}q) 
S_d(k_{-+}) \beta_d \g_{\mu}  \label{LambdaPid}
.
\ea
Here: $S_f(k)$ is the propagator for a quark of
flavor $f$;
$\Gamma_{\pi}(k)$ is the pion Bethe-Salpeter amplitude; 
$k_{\alpha \beta} = (k + \sfrac{\alpha}{2} q + \sfrac{\beta}{2} P)$; 
$N_c = 3$ is the number of colors;
and the trace is over Dirac indices. 
The kinematical variables for evaluating Eqs.~(\ref{LambdaPiu}) and
(\ref{LambdaPid}) are also indicated in \Fig{Fig:PiN}.  
The small mass difference between the $u$ and $d$ quarks is ignored
and hence we take $S_u(k) = S_d(k)$.
In this case, $\Lambda^u_{\mu}(q,P)$ and $\Lambda^{\bar d}_{\mu}(q,P)$ 
contribute equally to $\pi^+ N$ elastic scattering.  
 
In a general covariant gauge, the propagator for a quark of flavor
$f$ is 
\be
S_f(k) = - i \g \scdot k \; \sigma_V^f(k^2) + \sigma_S^f(k^2)
,
\label{QuarkPropDef}
\ee
and its inverse is
\be
        S^{-1}_f(k) = i \g \scdot k \; A_f(k^2) + B_f(k^2)
.
\label{QuarkPropDef2}
\ee

Studies of the 2-point quark DSE, employing a model-gluon
propagator, suggest that the qualitative features of the
confined-quark propagator are well described by the following
algebraic form \cite{FrankGluon}:  
\begin{eqnarray}
  \bar{\sigma}^f_S(x) &=& 
\frac{1\-e^{-b^f_1x}}{b^f_1 x} \frac{1\-e^{-b^f_3 x}}{b^f_3 x} 
\left(b^f_0 + b^f_2 \frac{1\-e^{-\Lambda x}}{\Lambda x} \right) 
\nn\ \\
& &   
+ \bar{m}_f\frac{1 \- e^{-2(x+\bar{m}_f^2)}}{x + \bar{m}^2_f }
   + C^f_{m_f} e^{- 2 x} , 
\nn  \\ 
\bar{\sigma}^f_V(x) &=& 
\frac{2(x+\bar{m}_f^2) - 1 + e^{-2(x+\bar{m}_f^2)} }
{ 2(x+\bar{m}_f^2)^2 },
   \label{QuarkProp}  
\end{eqnarray}
with $x = k^2/\lambda^2$, $\bar{\sigma}^f_{S} = \lambda \sigma^f_S$,
$\bar{\sigma}^f_V = \lambda^2 \sigma^f_V$, $m_f = \lambda \bar{m}_f$ 
is the bare mass for a quark of flavor $f$, $\Lambda = 10^{-4}$ 
and $\lambda = 0.566$ GeV is a momentum scale.
All of the parameters and variables appearing in \Eqs{QuarkProp} are
dimensionless as they have been rescaled by $\lambda$.
The value $\Lambda = 10^{-4}$ is introduced only to ensure the
decoupling of the small and intermediate $k^2$ behavior in the
algebraic form, characterized by the parameters $b_0^f$ and
$b_2^f$, which is observed in numerical studies of the quark DSE.

The parameters $b^f_0$,  $b^f_1$, $b^f_2 $, $b^f_3$, $C^f_{m_f}$, and
$m_f$ for $f = u,d,s$ are determined in Ref.~\cite{Burden} by
performing a $\chi^2$-fit to experimental values of decay constants
$f_\pi$, $f_K$, $\pi\pi$ scattering lengths, 
$\pi$- and $K$-meson electromagnetic form
factors, charge radii and other $\pi$- and $K$-meson observables.
The resulting values of the parameters are given in \Tab{Tab:QuarkParam}.

\begin{table}
\begin{center}
\begin{tabular}{l|llllllr}
$f$ & $b^f_0$ & $b^f_1$ & $b^f_2 $ & $b^f_3$ & $C^f_{m_f=0}$ &
$C^f_{m_f \neq 0} $ & $m_f \;\;\;\;\;\; $\\  \hline
$u,d$& 0.131   & 2.90   & 0.603  & 0.185 & 0.121 & 0.00 & 5.1 MeV\\
$ s $  & 0.105   & same   & 0.740  & same& 1.69  & same & 127.5 MeV
\end{tabular}
\caption{Confined-quark propagator parameters for $u$, $d$, and $s$
quarks from \Ref{Burden}. 
An entry of ``same'' for $s$ quark indicates that the value of the 
parameter is the same as $u$ and $d$ quarks.} 
\label{Tab:QuarkParam}
\end{center}
\end{table}

The algebraic forms for $\sigma_S^f(k^2)$ and $\sigma_V^f(k^2)$ are
{\em analytic} everywhere in the finite complex $k^2$-plane. 
This ensures that the quark propagator, $S_f(k)$, has no Lehmann 
representation and hence there are no quark-production thresholds in
the calculation of observables.  
The absence of such thresholds admits the interpretation
that $S_f(k)$ describes the propagation of a {\em confined} quark. 

It is possible that the {\em exact} quark propagator, obtained
directly from the quark DSE of QCD, may have singularities or 
branch cuts (perhaps both) somewhere in the complex-$k^2$ plane. 
In this case, the forms given in \Eqs{QuarkProp} are to be considered 
as approximate and only applicable within some region of Euclidean
space.  
In the present study, the quark propagator in \Eqs{QuarkProp} is
employed on a hyperbolic region of the complex Euclidean plane where 
${\rm Re}(k^2) \ge - \sfrac{1}{4} m^2_{\phi}$ and $m_{\phi}$ is the
mass of the $\phi$ meson.  
On this domain, our results indicate that the forms given in
\Eqs{QuarkProp} are sufficient to describe the hadron observables
considered herein. 
\begin{figure}[t]
\centering{
\epsfig{figure=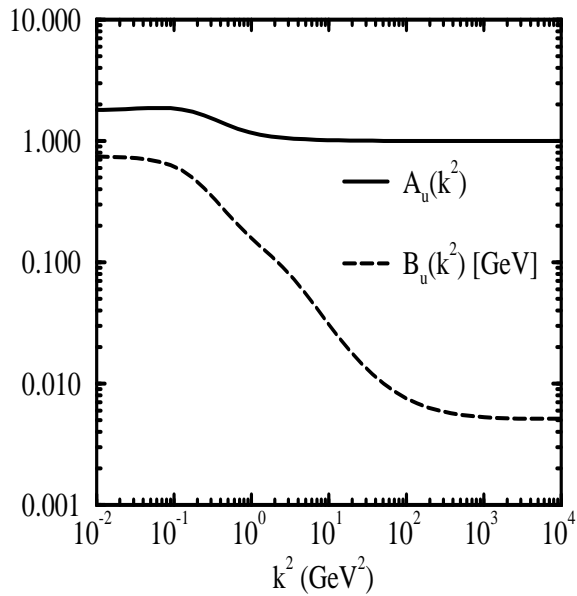,height=8.50cm,width=8.20cm}}
\caption{The Lorentz-invariant functions $A_u(k^2)$ and $B_u(k^2)$ for
the confined, $u$-quark propagator defined in \Eqs{QuarkProp}.}
\label{Fig:QuarkProp}
\end{figure}

The inverse propagator of a quark with flavor $f$ is defined in terms
of the two functions, $A_f(k^2)$ and $B_f(k^2)$, as given by
\Eq{QuarkPropDef2}. 
These are obtained from $\sigma_V^f(k^2)$ and $\sigma_S^f(k^2)$ by
inverting \Eq{QuarkPropDef}.
In \Fig{Fig:QuarkProp}, the functions $A_u(k^2)$ and $B_u(k^2)$ that
correspond to the confined $u$-quark propagator are shown in order to
illustrate some of their features, which play an important role in this
work. 

In the limit that the quark momentum, $k^2$, becomes large and
spacelike, the functions $A_f(k^2)$ and $B_f(k^2)$ approach an
asymptotic limit,
\ba
\lim_{k^2 \rightarrow \infty} \;\; A_f(k^2) &=& 1 , 
\label{ABpQCDLimit} \\
\lim_{k^2 \rightarrow \infty} \;\; B_f(k^2) &=& m_f  
.
\ea
Hence, the confined-quark propagator, $S_f(k)$, reduces to a 
free-fermion propagator, 
\be
\lim_{k^2 \rightarrow \infty} \;\; S_f^{-1}(k) = i\g\scdot k+m_f.
\label{FreeQuarkProp}
\ee
This behavior can be identified with the asymptotic freedom of the
quark propagator in QCD.  

For small quark momenta, the quark propagator, given by
\Eqs{QuarkProp}, behaves quite differently than the free-fermion
propagator of \Eq{FreeQuarkProp}.  
In this region, there is a strong nonperturbative enhancement of the
mass function, $B_f(k^2)$.
This enhancement is a manifestation of dynamical chiral symmetry
breaking and confinement.  

Dynamical chiral symmetry breaking causes the low-momentum quark
propagator to behave qualitatively like a constituent-quark
propagator.   
(A constituent-quark propagator is given by the free-fermion
propagator of \Eq{FreeQuarkProp}, but with a mass of the order of a
hadron mass.)   
One can identify an {\em effective} constituent-quark mass\footnote 
{The reader is reminded that a confined-quark propagator does not have
a well-defined mass.  Hence, the procedure employed to define an
``effective mass'' is arbitrary.}
for the confined-quark propagator given in \Eqs{QuarkProp}
as the solution of $k^2 - B_f(k^2) / A_f(k^2) = 0$. 
With this definition, we find the effective $u$-quark mass is 
$M^u_Q \approx$ 330~MeV and effective $s$-quark mass is 
$M^s_Q \approx$ 490~MeV. These masses are typical of the values
employed in constituent-quark models.  

The fact that the quark propagator of \Eqs{QuarkProp} behaves like a 
constituent-quark propagator at low momentum and a current-quark
propagator at high momentum has important ramifications in our
analysis of vector-meson electroproduction in \Sec{Sec:Vector}.

We now consider the Bethe-Salpeter amplitude, $\Gamma_\pi(k)$.
In the limit, $m_u \rightarrow 0$, the $\pi$ meson can be identified
with the Goldstone mode of massless QCD and one can show that the
corresponding BS equation coincides with DSE for the quark
mass function, $B_u(k^2)$ \cite{Bender}.    
Therefore, the pion BS amplitude can be written as 
\be
\Gamma_{\pi}(k)  =  i \g_5 \left. \frac{B_u(k^2)}{f_{\pi}} 
\right|_{m_u = 0}
\label{PiBSAmp},
\ee
where $B_u(k^2)$ is given by \Eq{QuarkPropDef2} and $f_{\pi}$ is the
BS normalization. 

Explicit chiral symmetry breaking effects associated with finite
current-quark mass are provided for by allowing $C^u_{m_u=0} \not = 0$
in \Tab{Tab:QuarkParam}, then the small value of $C^u_{m_u=0}$
obtained in \Ref{Burden} and given in \Tab{Tab:QuarkParam} suggests
that the $\pi$ meson is, to a good approximation, a Goldstone mode of
QCD\@. 

\begin{figure}
{\centering\ 
\epsfig{figure=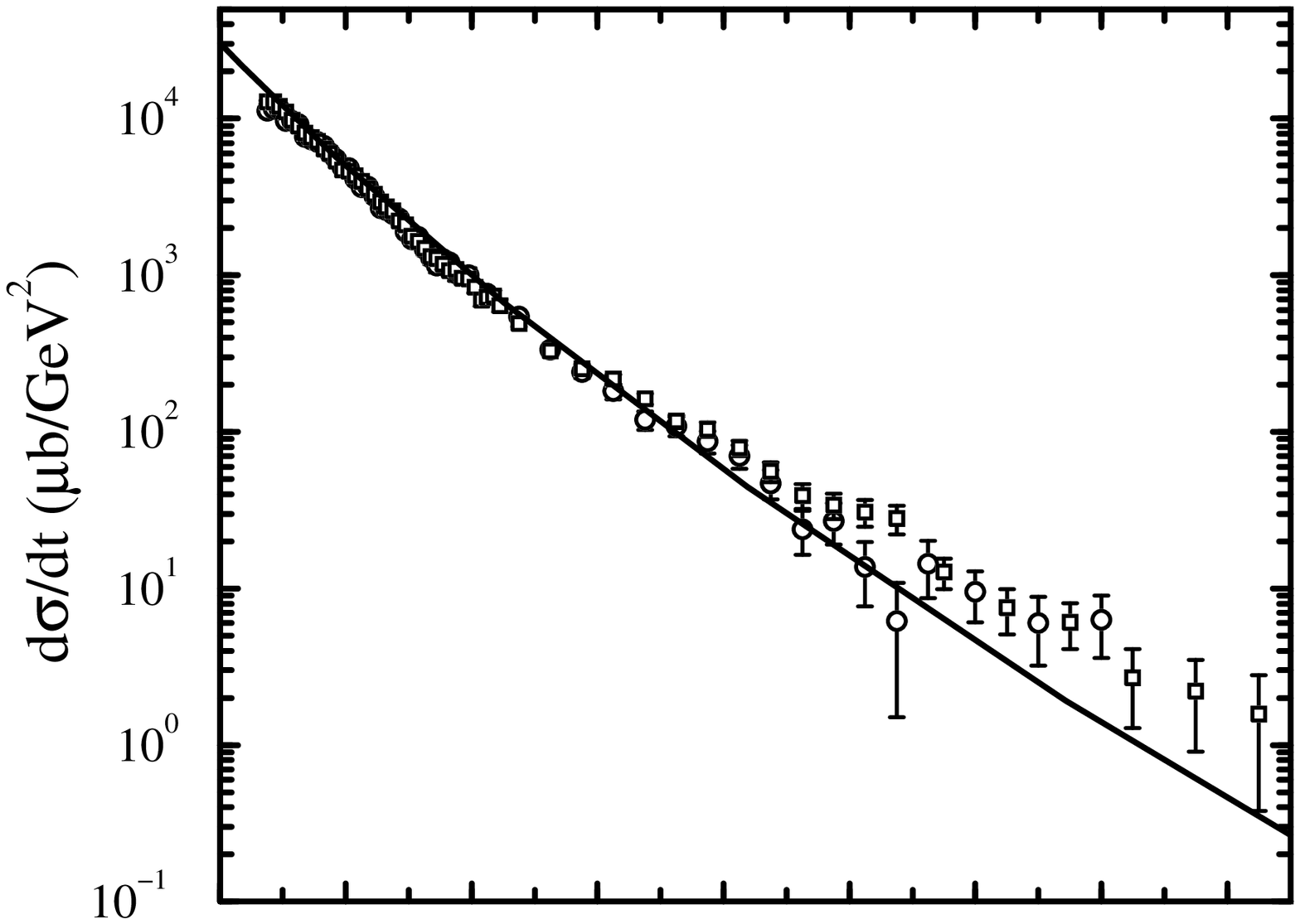,height=8.0cm,width=8.0cm}
}\vspace*{-2.00cm}

{\centering\ 
\epsfig{figure=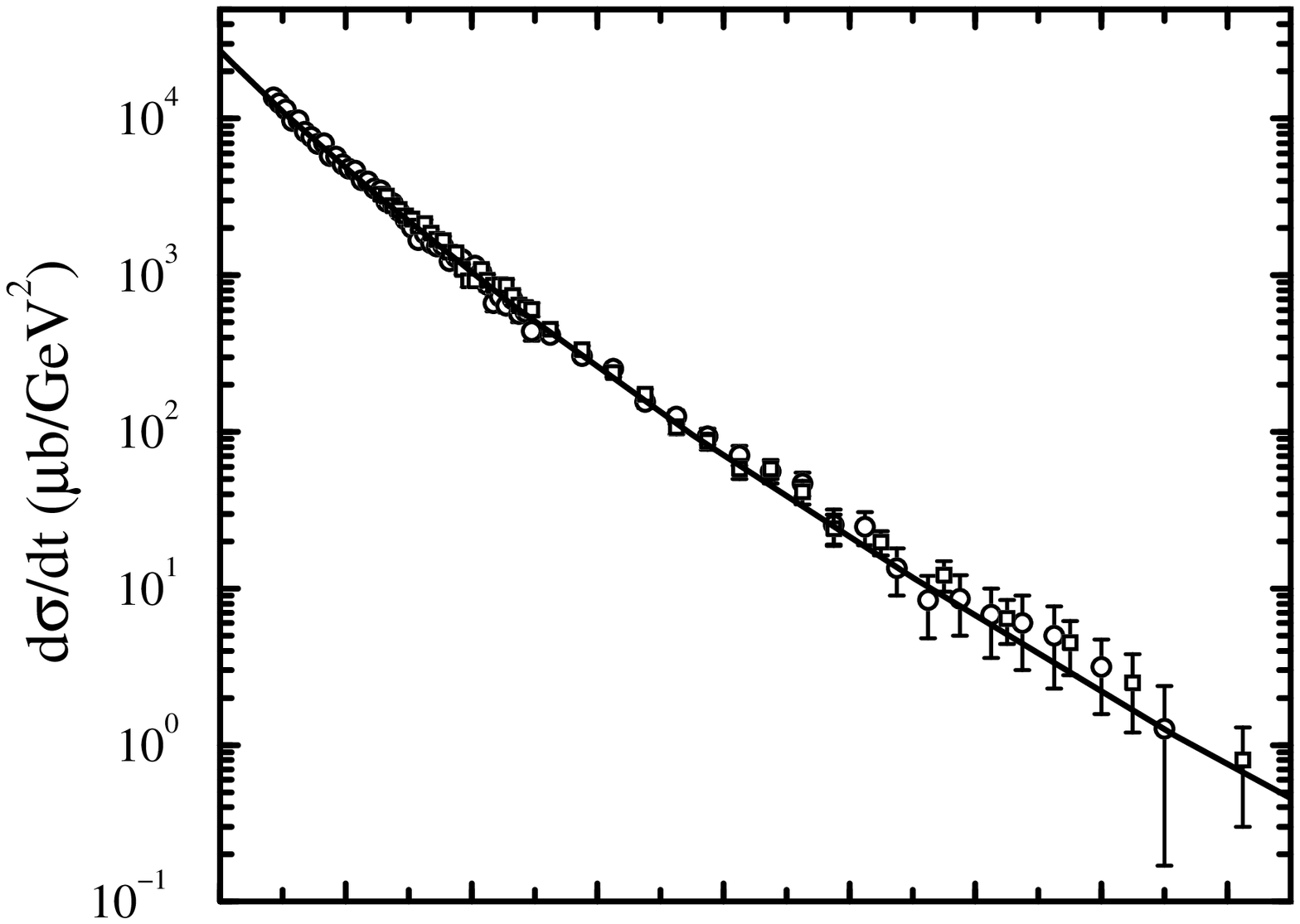,height=8.0cm,width=8.0cm}
}\vspace*{-2.0cm}

{\centering\ 
\epsfig{figure=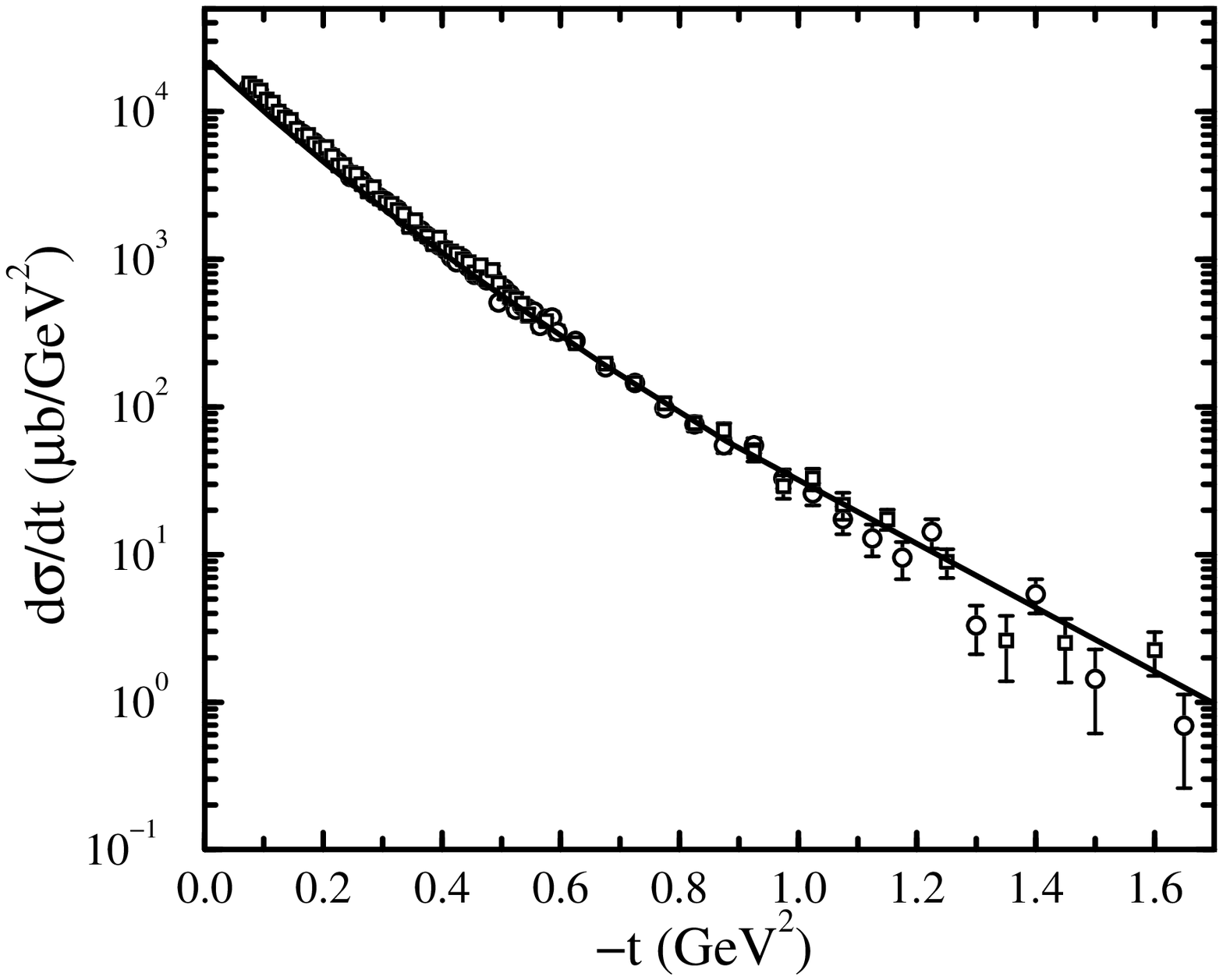,height=8.0cm,width=8.0cm}
}
\caption{
The differential cross section for $\pi p$ elastic scattering at
CM energies $\protect\sqrt{s} = $ 20 (top), 15 (middle) and 10~GeV
(bottom).  The data are from \Ref{AkerlofElastic}. 
Both $\pi^+ p$ (circles) and $\pi^- p$ (squares) data are shown.
}
\label{Fig:Pion}
\end{figure}
%

Using the confined-quark propagator, $S_f(k)$ with \Eqs{QuarkProp}, 
$\pi$-meson BS amplitude, $\Gamma_{\pi}(k)$ in
\Eq{PiBSAmp}, the quark-loop integrations of Eqs.~(\ref{LambdaPiu})
and (\ref{LambdaPid}) are evaluated numerically.
The $\pi N$ elastic scattering differential cross section is then
obtained from 
\be
\frac{d \sigma}{dt} = \frac{M_N^2}{16 \pi s |\vec{p}_{\rm cm} |^2}
\frac{1}{2} \sum_{m_s,m_s^\prime} | T_{\pi N \rightarrow \pi N} |^2
\label{Defdsigdt}
.
\ee
The parameters $\beta_u$ and $\alpha_1$ are chosen to provide the best 
agreement with existing $\pi N$ elastic scattering data in the
high-energy and small-momentum-transfer region. 
The resulting parameters are given in the first row of
\Tab{Tab:PomParam}.  
The results (solid curves) and the data at $\sqrt{s}=$ 10, 15 and
20~GeV are shown in \Fig{Fig:Pion}. 
Clearly, the model provides a good description of the $\pi N$
differential cross section.

\begin{table}[t]
\begin{center}
\begin{tabular}{l|lll} 
flavor&$\alpha_0$&$\alpha_1$ [GeV$^{-2}$]&$\beta_f$ [GeV$^{-2}$]
\\ \hline
$u/d$ & 0.10 & 0.33 & 2.35 \\
$s$ & same & same & 1.50 \\
$c$ & 0.42 & 0.10 & 0.09 
\end{tabular}
\caption{{\sloppypar 
Parameters of the quark-nucleon Pomeron-exchange interaction
of \Eq{PomDef}. Entry of ``same'' indicates that value is the same as for
$u$- and $d$-quarks.
}}
\label{Tab:PomParam}
\end{center}
\end{table}

To determine the remaining parameter, $\beta_s$, we apply our
model interaction to $K N$ elastic scattering.      
The $K^+ N$ elastic scattering amplitude can be obtained from
\Eq{TPiN} by making the substitution  $\pi \rightarrow K$ and using
Eqs.~(\ref{LambdaPiu}) and (\ref{LambdaPid}) with the substitutions
$\pi \rightarrow K$ and $d \rightarrow s$.   
The amplitude can be  calculated immediately by employing the
$s$-quark propagator of \Eqs{QuarkProp},  
the $s$-quark parameters given in \Tab{Tab:QuarkParam} and the
$K$-meson BS amplitude, $\Gamma_K(k)$, determined from \Eq{PiBSAmp}
with the replacements $\pi \rightarrow K$ and $u \rightarrow s$.

\begin{figure}
{\centering\ 
\epsfig{figure=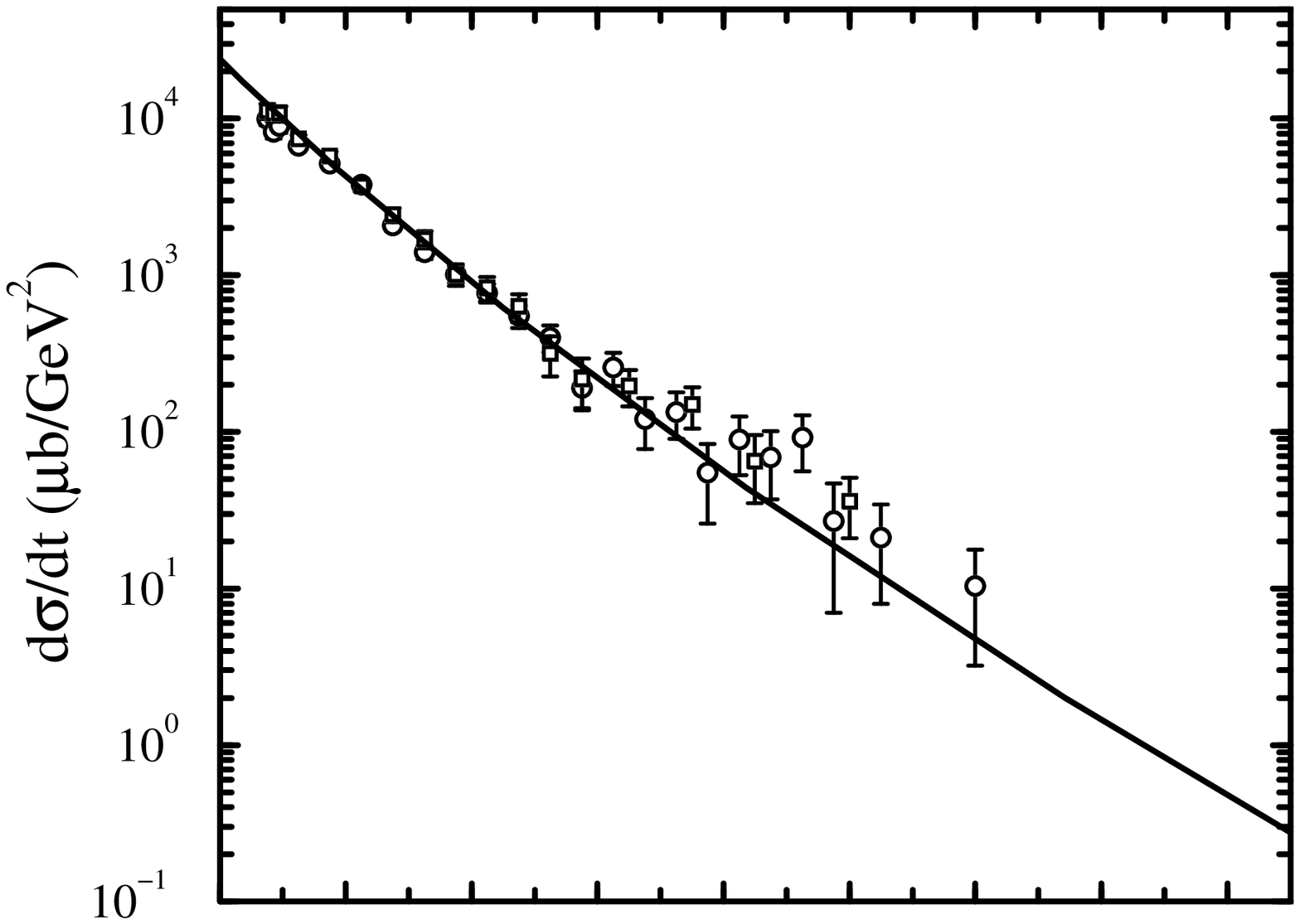,height=8.0cm,width=8.0cm}
}\vspace*{-2.0cm}

{\centering\ 
\epsfig{figure=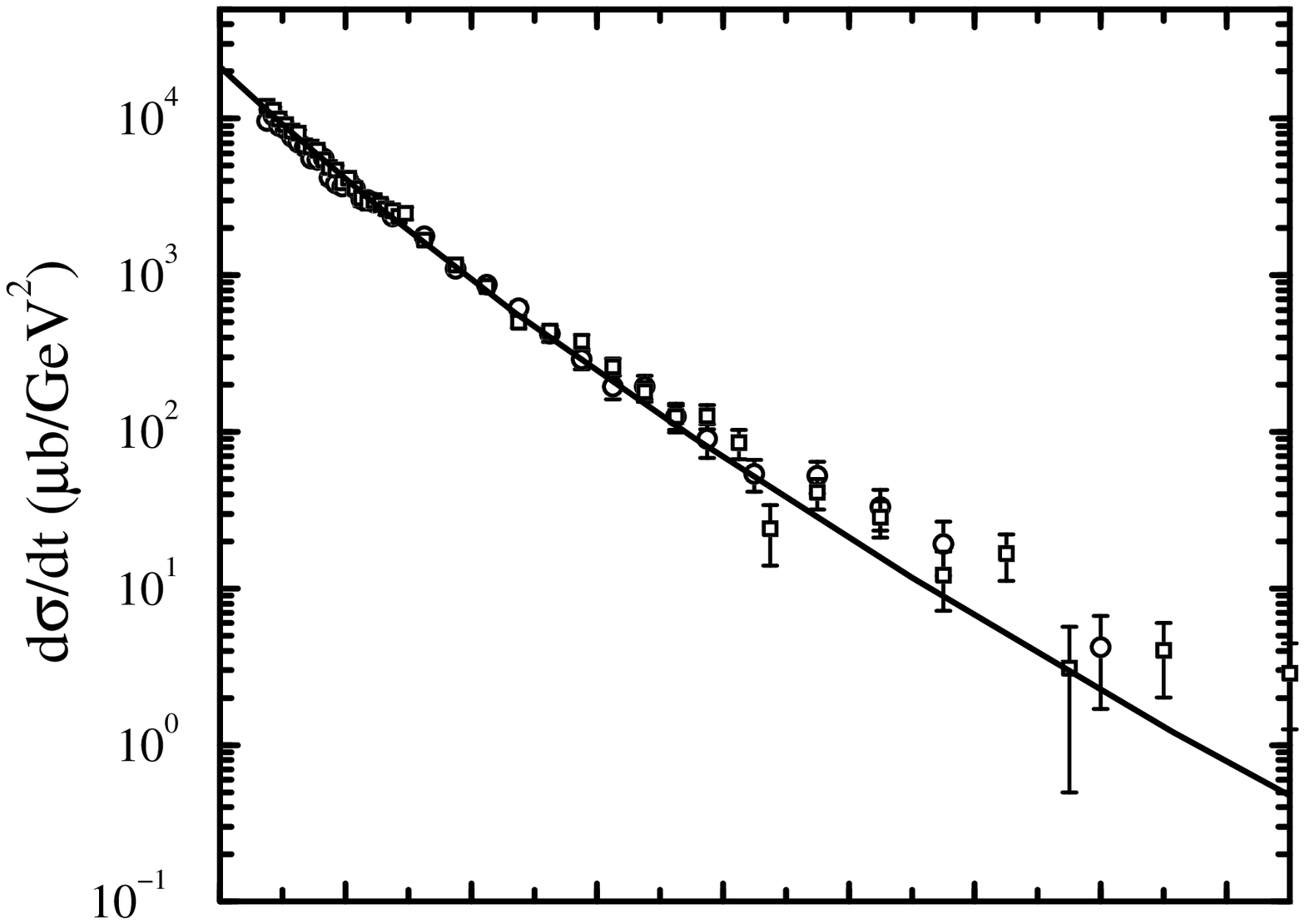,height=8.0cm,width=8.0cm}
}\vspace*{-2.0cm}

{\centering\
\epsfig{figure=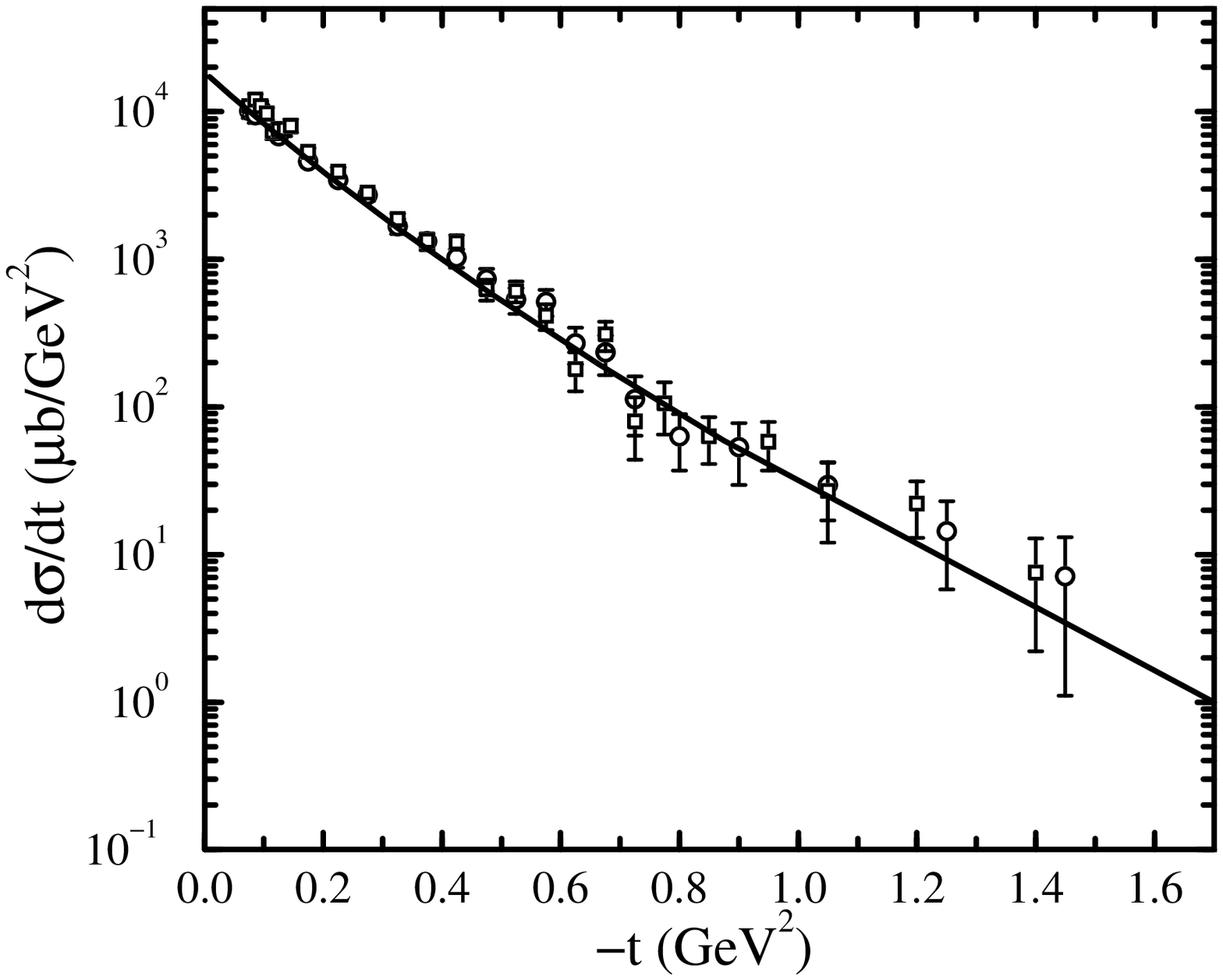,height=8.0cm,width=8.0cm}
}
\vspace*{1.0cm}
\caption{
The differential cross section for $K p$ elastic scattering at CM
energies $\protect\sqrt{s} =$ 20 (top), 15 (middle) and 10~GeV
(bottom). 
The data are from \Ref{AkerlofElastic}. Both $K^+ p$ (circles)
and $K^- p$ (squares) data are shown.
}
\label{Fig:Kaon}
\end{figure}

We find that the {\em magnitude} of the $K N$ elastic scattering data
can be  best reproduced by choosing $\beta_s = 1.5$  GeV$^{-1}$, as
listed in \Tab{Tab:PomParam}. 
Our results for $\sqrt{s}= 10, 15$ and 20~GeV are shown
in \Fig{Fig:Kaon}.  The agreement with the data is excellent. 
We emphasize that the $t$ dependence of the differential cross section 
shown in \Fig{Fig:Kaon} is a {\em prediction} of the model since   
all of the $s$ and $t$ dependence of the quark-nucleon Pomeron-exchange
interaction has been fixed by the $\pi N$ elastic scattering data. 
Our results for the differential cross section, shown in
\Figs{Fig:Pion} and \ref{Fig:Kaon}, are well parameterized by the
following form: 
\be
\frac{d\sigma}{dt} = A e^{b t},
\ee
where $A$ and $b$ are functions of $s$ only. 
A determination of $b_{\pi N}$ and $b_{KN}$ from our model reveals
that although our quark-nucleon Pomeron-exchange interaction,
$G(s,t)$ in \Eq{GDef}, has the same $t$ dependence for both of these
processes, nonetheless, because of the quark-loop integration in
\Eqs{LambdaPiu} and (\ref{LambdaPid}), $b_{\pi N} \not= b_{KN}$.  
In particular, we find that at $\sqrt{s} = 15$ GeV, 
$b_{\pi N} = 9.17$ GeV$^{-2}$ and $b_{KN} = 8.82$ GeV$^{-2}$. 
This flavor dependence of the $t$-slope parameter arises from the
dynamical difference between the quark substructure of the $\pi$ and
$K$ meson.  
The only other source of flavor dependence in our approach 
is the coupling {\em constant}, $\beta_f$, which is independent of $s$
and $t$.  
This illustrates how flavor dependence in diffractive processes can
arise from the quark substructure of the hadrons involved.  In the
next section, we show that the different quark substructure of the
$\rho$, $\phi$ and $J/\psi$ mesons leads to striking differences in the
$q^2$ dependence of their electroproduction cross sections.

Having defined all of the parameters in the quark-nucleon
Pomeron-exchange interaction of \Eqs{GDef} and (\ref{PomDef}), we are
now in a position to make parameter-free predictions for other
diffractive processes on a nucleon target. 

\section{Diffractive electroproduction of vector mesons}
\label{Sec:Vector}
Motivated by recent experimental efforts at HERA, FNAL and TJNAF, 
the remainder of this paper focuses on the diffractive
electroproduction of vector mesons on the nucleon. 
Our main objective is to explore the role played by the quark
substructure of vector mesons in diffractive electroproduction.
We also examine the extent to which the quark-nucleon Pomeron-exchange 
interaction defined by \Eq{PomDef} can account for the existing data
and make predictions for future experiments.
 
To proceed, it is necessary to develop model quark-antiquark 
BS amplitudes for the vector mesons under consideration. 
In principle, one could solve the BS equation directly to obtain these
amplitudes.  
However, for our present purposes, it is sufficient to 
construct phenomenological amplitudes that reproduce the experimental
decay widths of the vector mesons.   
In what follows, only formulae necessary to the calculation of $\rho$
decays are given. 
By making an appropriate change of labels, the same formulae can be
applied to calculate the decays of $\phi$ and $J/\psi$ mesons.

The vector-meson BS amplitudes obtained in numerical DSE
studies of the meson spectrum indicate that the qualitative features
of these amplitudes are finite at $k^2=0$ and evolve as $1/k^2$ for 
large $k^2$ \cite{Jain}.  
Therefore, we parameterize the amplitudes,
\be
V_{\mu}(p,p^\prime) = \frac{1}{N_V} 
\left( e^{-k^2/a_V^2} + \frac{c_V}{1\+k^2/b^2_V} \right) 
\left[\gamma_{\mu}+\frac{P_{\mu}\gamma\cdot P}{m_V^2} 
\right],
\label{VectorBSAmp}
\ee
where $P=p \- p^{\prime}$ and $m_V$ are the total momentum and mass of
the vector meson and $k = \sfrac{1}{2}(p+p^\prime)$ is the relative
momentum between the quark and antiquark. 
The parameters $a_V$, $b_V$ and $c_V$ represent, respectively, the
infrared scale, the scale of the $1/k^2$ tail and their
relative strength.  The overall normalization, $N_V$, is {\em not} a
free parameter. It is determined by the BS normalization equation, 
\ba
\lefteqn{P_{\alpha} 
\left(\delta_{\mu\nu}\+\frac{P_{\mu}P_{\nu}}{m_{\rho}^2}\right)
=} \nn\ \\
& & N_c  {\rm tr} \intd{k} 
\frac{ \partial S_u(k_+)  }{ \partial P_{\alpha} }
V_{\mu}(k_+,k_-) S_u(k_-) V_{\nu}(k_-,k_+)
 \label{VectorBSN}
\\
& &+  N_c  {\rm tr} \intd{k} S_u(k_+) V_{\mu}(k_+,k_-) 
\frac{ \partial S_u(k_-)  }{ \partial P_{\alpha} } V_{\nu}(k_-,k_+)
\nn\ .
\ea

The Dirac structure of the BS amplitude in \Eq{VectorBSAmp} is not the 
most general form. Rather, it is the simplest form to ensure that 
the {\em Lorentz condition} is satisfied by the on-mass-shell vector
meson; i.e.\ $P_{\mu} V_{\mu}(p,p^\prime) = 0$.

\begin{figure}[t]
\centering{\ 
\mbox{\epsfig{figure=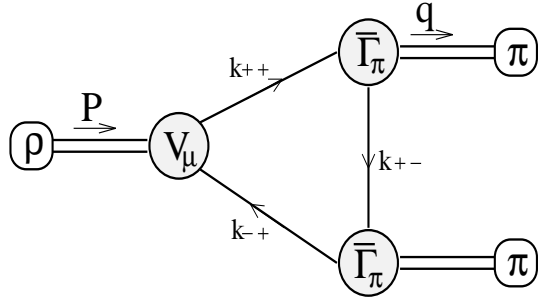,height=4.0cm,width=7.0cm}}}
\caption{Impulse approximation to $\rho \rightarrow \pi \pi$ decay.}
\label{Fig:RhoPiPi}
\end{figure}

To determine the parameters in $V_{\mu}(p,p^\prime)$, we first 
consider the decay of a neutral-$\rho$ meson of momentum $P$ 
into a $\pi^+$-meson of momentum $q$ and $\pi^-$-meson of momentum 
$q^\prime = P-q$. In impulse approximation, the 
$\rho \rightarrow \pi \pi$ transition amplitude is  
\be
\langle q;P\-q | T_{\rho\rightarrow\pi\pi} | P \lambda_\rho \rangle 
= - 2 \Lambda_{\mu}(P,q) \varepsilon_{\mu}^{\lambda_{\rho}}(P)
\label{TRhoPiPi}
,
\ee
where $\varepsilon_{\mu}^{\lambda_{\rho}}(P)$ is the 
polarization vector of a $\rho$-meson with helicity $\lambda_{\rho}$
and 
\ba
\Lambda_{\mu}(P,q) &=& 2
N_c {\rm tr}\intd{k} 
S_u(k_{-+}) V_{\mu}(k_{-+},k_{++}) 
\nn\ \\ & & \times
S_u(k_{++}) \bar{\Gamma}_{\pi}(k + \sfrac{1}{2} P)
S_d(k_{+-}) \bar{\Gamma}_{\pi}(k)
. \label{RhoPiPi}
\ea
Here, $k_{\alpha \beta} = k + \sfrac{\alpha}{2} P 
+ \sfrac{\beta}{2} q$,
$N_c = 3$ and the trace is over Dirac indices.
The Feynman diagram corresponding to \Eq{RhoPiPi} is depicted in
\Fig{Fig:RhoPiPi}.  
Since the quark propagator, $S_f(k)$, and $\pi$-meson BS 
amplitude, $\Gamma_{\pi}(k)$, have been given in the previous section,
the only new element in \Eq{RhoPiPi} is the $\rho$-meson
BS amplitude, $V_{\mu}(p,p^{\prime})$.   
Therefore, \Eq{RhoPiPi} provides a means to
determine the parameters, $a_{\rho}$, $b_{\rho}$ and $c_{\rho}$,
in the $\rho$-meson BS amplitude, $V_{\mu}(p,p^\prime)$.

We define the $\rho \rightarrow \pi \pi$ decay constant,
$g_{\rho\pi\pi}$, by writing \Eq{RhoPiPi} as
\be
\Lambda_{\mu}(P,q) = \frac{1}{4} (2q_{\mu} - P_{\mu}) g_{\rho \pi \pi}
        \Lambda(P^2,q^2)
\label{GRhoPiPi}
,
\ee
with 
$\Lambda(P^2 = - m_{\rho}^2,q^2 = - m^2_{\pi}) = 1$ 
when all external legs of \Fig{Fig:RhoPiPi} are on-mass-shell. 
The $\rho \rightarrow \pi \pi$ decay width, 
$\Gamma_{\rho \rightarrow \pi \pi}$, 
is related to this decay constant by the following expression:
\be
\Gamma_{\rho \rightarrow \pi \pi} = \frac{ g_{\rho \pi \pi}^2}{4 \pi}
\frac{m_{\rho}}{12} \left( 1 - \frac{4 m_{\pi}^2}{m_{\rho}^2} 
\right)^{3/2}
.
\label{GRhoPiPiDef}
\ee
Substituting the experimental value of 
$\Gamma_{\rho \rightarrow \pi\pi} = 151.2 \pm$ 1.2~MeV \cite {PDG} 
into \Eq{GRhoPiPiDef} leads to $g_{\rho \pi\pi} = 6.05$.  
Upon substituting the explicit form for $V_{\mu}(p,p^{\prime})$
into \Eq{RhoPiPi}, we determine the value of the decay
constant $g_{\rho \pi \pi}$ in our model. 

\begin{figure}[t]
\centering{\ 
\mbox{\epsfig{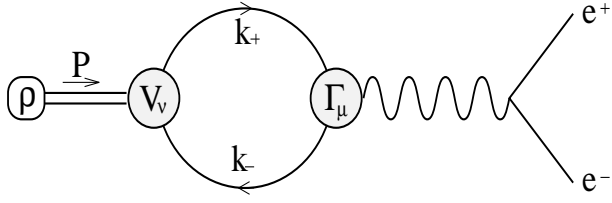}}}
\caption{Impulse approximation to $\rho \rightarrow e^+ e^-$ decay.}
\label{Fig:RhoGee}
\end{figure}

We also consider the $\rho$-meson electromagnetic decay, 
$\rho \rightarrow e^+ e^-$, which proceeds through an intermediate 
timelike photon.
The transition amplitude for the decay of a $\rho$-meson with 
momentum $P$ into an electron of momentum $p$ and a positron of
momentum $p^{\prime}$ takes the following form:
\ba
\lefteqn{
\langle p m_s ; p^{\prime} m^{\prime}_s | 
T_{\rho\rightarrow e^+ e^-}
| P \lambda_{\rho} \rangle  =
}
\nn\ \\ & & e_0 \bar{u}_{e}(p)  \gamma_{\mu} v_{e}(p^{\prime}) 
\frac{1}{P^2}  
\Pi_{\mu \nu}(P)  \varepsilon_{\nu}^{\lambda_{\rho}}(P)
,
\ea
where $\bar{u}_{e}(p)$ and $v_e(p^\prime)$ are respectively the
spinors of  
the outgoing electron and positron, $e_0$ is the positron charge, 
$\Pi_{\mu\nu}(P)$ is the photon-$\rho$-meson transition amplitude and
$\varepsilon_{\nu}^{\lambda_\rho}(p)$
is the $\rho$-meson polarization vector. 
In impulse approximation, the photon-$\rho$-meson transition
amplitude is given by
\ba
\Pi_{\mu \nu}(P) &=& e_0 N_c {\rm tr} \intd{k}
S_u(k_+) \nn\ \\
& & \times \Gamma_{\mu}(k_+,k_-) S_u(k_-) V_{\nu}(k_-,k_+), 
\label{RhoG}
\ea
and is illustrated in \Fig{Fig:RhoGee}.

Evaluation of \Eq{RhoG} requires an explicit form for the photon-quark
vertex, $\Gamma_{\mu}(k,k^\prime)$. An acceptable form must satisfy the
Ward-Takahashi identity (WTI), 
\be
(k - k^{\prime})_{\mu} 
i \Gamma_{\mu}(k,k^\prime) = S_f^{-1}(k) - S_f^{-1}(k^{\prime})
,
\label{WTI}
\ee
where $S_f(k)$ is the quark propagator.
Clearly, a bare quark-photon vertex, $\Gamma_{\mu} = \gamma_{\mu}$,
does not satisfy \Eq{WTI} for the confined-quark propagators of
\Eqs{QuarkProp}.  

The most general quark-photon vertex that satisfies \Eq{WTI} and is
free of kinematic singularities is given by \cite{Bashir}
\be
        i \Gamma_{\mu}(k,k^\prime) = 
        i \Gamma_{\mu}^{\rm T}(k,k^\prime) + 
        i \Gamma_{\mu}^{\rm BC}(k,k^\prime)
,
\ee
where
\be
       (k-k^\prime)_{\mu}  \Gamma_{\mu}^{\rm T}(k,k^\prime)  = 0
,
\ee
and \cite{Ball}
\ba
\lefteqn{i \Gamma_{\mu}^{\rm BC}(k,k^{\prime}) = 
 i \g_{\mu} f_1(k^2,k^{\prime 2}) }
\nn\ \\
& & + (k+k^{\prime})_{\mu} [i \g \!\cdot \sfrac{k+k^{\prime}}{2}
f_2(k^2,k^{\prime 2}) + f_3(k^2,k^{\prime 2})],
\label{GammaBC}
\ea
with
\ba
f_1(k^2,k^{\prime 2}) &=& \frac{A_f(k^2) + A_f(k^{\prime 2})}{2}
, \nn \\ 
f_2(k^2,k^{\prime 2}) &=& \frac{A_f(k^2) - A_f(k^{\prime 2})}
	{k^2 - k^{\prime 2}}
, \label{GammaBCf} \\
f_3(k^2,k^{\prime 2}) &=& \frac{B_f(k^2) - B_f(k^{\prime 2})}
	{k^2 - k^{\prime 2}}
.  \nn
\ea
The functions $A_f(k^2)$ and $B_f(k^2)$ are the Lorentz invariant
functions that determine the inverse of the confined-quark propagator
in \Eq{QuarkPropDef2}. 

We observe from \Eqs{GammaBC} and (\ref{GammaBCf}) that 
$\Gamma_{\mu}^{\rm BC}(k,k^\prime)$ reduces to a bare quark-photon
vertex, $\gamma_{\mu}$, in the limit that both quark momenta become
large and spacelike, in accordance with perturbative QCD. 
Hence, in this perturbative limit, $\Gamma_{\mu}^{\rm T}(k,k^\prime)$
must vanish.   
Of course, at $q = k - k^{\prime} = 0$ the Ward identity fixes the full
quark-photon 
vertex to be equal to $\Gamma_{\mu}^{\rm BC}(k,k)$, therefore, in this
limit $\Gamma^{\rm T}_{\mu}(k,k)$ must also vanish. 
Furthermore, numerical studies indicate that in the spacelike region, 
$\Gamma_{\mu}^{\rm T}(k,k^\prime)$ is a slowly varying function of
photon momentum. 
With these considerations, it is reasonable to neglect contributions
to the quark-photon vertex from $\Gamma_{\mu}^{\rm
T}(k,k^\prime)$. So, 
throughout this work we use 
\be
\Gamma_{\mu}(k,k^\prime) = \Gamma^{\rm BC}_{\mu}(k,k^\prime)
.
\ee
This ensures that we have a parameter-free quark-photon
vertex  that satisfies the WTI and has the correct
perturbative limit as both $k^2$ and $k^{\prime 2}$ become large and
spacelike. Furthermore, it has the correct transformation properties
under $C$, $P$, $T$ and Lorentz transformations. 

Since the $\rho$-meson is on-mass-shell,
$\Pi_{\mu \nu}(P)$ is transverse to
$P_{\mu}$. We can, therefore, define the dimensionless $\rho$-meson 
electromagnetic decay constant, $f_{\rho}$, by 
\be
 \Pi_{\mu \nu}(P) \left|_{P^2 = - m^2_{\rho}}
= \left( P^2 \delta_{\mu \nu} + P_{\mu} P_{\nu}
\right)  \frac{e_0}{f_{\rho}} \right|_{P^2=-m^2_{\rho}}.
\label{FRhoDef}
\ee
One can then show that the decay width for $\rho \rightarrow e^+ e^-$
is given by 
\be
\Gamma_{\rho \rightarrow e^+ e^-} = \frac{1}{3} \alpha_{\rm em}^2
m_{\rho}  \frac{ 4 \pi}{f^2_{\rho}}
\label{RhoeeWidth}
,
\ee
where $\alpha_{\rm em} \equiv e_0^2/ 4 \pi \approx 1/137 $ 
is the fine structure constant of QED. 
The experimental value of $\Gamma_{\rho\rightarrow e^+ e^-}=6.77\pm$ 
0.32~keV  \cite{PDG} entails $f_{\rho} = 5.03$. 

With the above formulae, we can use the experimental data of 
the $\rho \rightarrow \pi \pi$ and $\rho \rightarrow e^+e^-$ decay
widths to constrain the parameters of the $\rho$-meson BS amplitude,
$V_{\mu}(p,p^\prime)$.
The parameters $a_\rho$, $b_\rho$, and $c_\rho$ are chosen to 
reproduce the experimental values of the decay constants, $f_{\rho}$
and $g_{\rho \pi\pi}$, to within 15\% of the experimental values. 
This accuracy is easily acheived with our simple model of the
vector-meson BS amplitude. Furthermore, our results for the $W$, $t$
and $q^2$ dependence of vector-meson electroproduction observables are
completely insensitive to changes made to the parameters of the BS
amplitude. 
Only the magnitudes of decay constants and the electroproduction cross
section are sensitive to changes in the model BS amplitude.
The chosen parameters for the $\rho$-meson BS amplitude and the
resulting decay constants are compared to the experimental values for
the decay constants in the first row of Table~\ref{Tab:VectorParam}.

Relabeling $\rho \rightarrow \phi$ and $u \rightarrow s$ in 
\Eqs{RhoG} and (\ref{RhoeeWidth}) and $\rho \rightarrow \phi$, $\pi
\rightarrow K$ and $u \rightarrow s$ in \Eqs{RhoPiPi},
(\ref{GRhoPiPi}) and (\ref{GRhoPiPiDef})  and multiplying \Eq{RhoG} by
$\sqrt{\sfrac{2}{9}}$ to take account of the flavor structure of the
$\phi$ meson, one can determine the BS amplitude for the $\phi$-meson
by reproducing the decay constants for $\phi \rightarrow \bar{K}K$ and
$\phi \rightarrow e^+e^-$.  
The resulting parameters and decay constants are given in
\Tab{Tab:VectorParam}. 

Having completed the determination of the necessary BS amplitudes, 
we can now predict diffractive, $\rho$- and $\phi$-meson
electroproduction cross sections.  This is carried out in the next
sections. 

\begin{table}
\begin{center}
\begin{tabular}{ l | rrr | cc } 
$V$-meson & $a_V$ (GeV) & $b_V$ (GeV) & $c_V$ 
\hspace*{0.2cm} & $f_V$ & $g_{VPP}$ \\ 
\hline
$\rho$ & 0.400 & 0.008 & 125.0 & 4.55 (5.03) &  6.8 (6.05) \\
$\phi$ & 0.450 & 0.400 &   0.5 & 14.8 (12.9) & 3.7 (4.55) \\
$J/\psi$ &  1.200 & --- & 0.0 & 11.5 (11.5) & --- (---)  
\end{tabular}
\caption{Parameters of the model vector-meson BS amplitudes and their 
corresponding decay constants. 
$VPP$ refers to the vector-meson decay into two pseudoscalar mesons, 
e.g. $\rho \rightarrow \pi \pi$ and $\phi \rightarrow K \bar{K}$.
The experimental values are given in parentheses.}
\label{Tab:VectorParam}
\end{center}
\end{table}

\subsection{Electroproduction of $\rho$-mesons}
\label{Sec:VectorRho}
The fact that a neutral vector meson and photon are both 
$J^{PC} = 1^{--}$ states suggests that diffractive vector-meson
electroproduction would be similar to elastic hadron scattering, in
that both processes satisfy the empirical rule of Gribov and Morrison
(see, for example, \Ref{Collins}). 
We therefore expect that diffractive electroproduction of neutral
vector mesons would be well described within our Pomeron-exchange
model.  
The first application of a quark-based Pomeron-exchange model to the
study of vector-meson electroproduction is given in \Ref{DL87}.

We argue that of the processes considered herein, vector-meson
electroproduction is {\em most} sensitive to the quark substructure of
mesons and a proper treatment of the confined-quark dynamics is
essential to obtain a description of vector-meson electroproduction. 
For example, when we neglect some of the nonperturbative aspects of
quark propagation, such as confinement and dynamical chiral symmetry
breaking, we find that agreement with electroproduction data requires
the inclusion of a quark-Pomeron-exchange form factor \cite{Pichowsky}.
In contrast to this, herein, we observe that the {\em same}
quark-nucleon Pomeron-exchange interaction that describes $\pi N$ and
$KN$ elastic scattering also provides an excellent description of
$\rho$- and $\phi$-meson electroproduction without changing any
parameters and without introducing a quark-Pomeron form factor.

The differential cross section for vector-meson electroproduction 
can be written as
\be
 \frac{ d \sigma}{dE_{e^\prime} d\Omega_{e^\prime} } 
= \Gamma \left(  \sigma_T + \epsilon \sigma_L \right)
.
\label{SigmaTSigmaL}
\ee
Here, we have introduced the lepton kinematical factors, $\Gamma$ and 
$\epsilon$, as in \Ref{Akerlof}.
The definitions of $\sigma_T$ and $\sigma_L$ admit their
identification as the {\em transverse} and {\em longitudinal} cross
sections of the {\em virtual} process:  
$\gamma^* N \rightarrow \rho^0 N$. 

The differential forms of $\sigma_T$ and $\sigma_L$, in the
center-of-momentum (CM) frame of the $\rho N$ system, can be written
as 
\ba
\frac{d\sigma_T}{d\Omega} &=&  \frac{1}{(2\pi)^2} \frac{M_N}{4 W}
\frac{|\vec{P}|}{K_H}  \frac{1}{2} \sum_{\rm spins}
(J_xJ_x^{\dag}+J_y J_y^{\dag})
, \nonumber \\
\frac{d\sigma_L}{d\Omega} &=& \frac{1}{(2\pi)^2} \frac{M_N}{4 W}
\frac{|\vec{P}|}{K_H}  \frac{q^2}{\omega^2}
\sum_{\rm spins}  (J_zJ_z^{\dag} ).
\ea
Here, $d\Omega$ denotes a differential element of 
the solid angle of the outgoing $\rho$-meson 3-momentum, $\vec{P}$, 
relative to the photon 3-momentum $\vec{q}$, 
$K_H = (W^2 -M_N^2) / (2 M_N)$,  
$ W^2 = - (P + p_2)^2$ is the invariant mass of the $\rho N$ system,  
$q_{\mu} = ( \vec{q}, i \omega)$ is the momentum of virtual photon
and $P_{\mu} = (\vec{P},i \sqrt{m_{\rho}^2 + |\vec{P}|^2})$ is
momentum of the outgoing $\rho$ meson. 
The hadron current matrix element is 
$J_{\mu} = \langle P \lambda_\rho ;p_2 m^{\prime} | 
J_{\mu}(q) | p_1 m \rangle $. 
\begin{figure}[t]
\centering{\ 
\mbox{\ 
\epsfig{figure=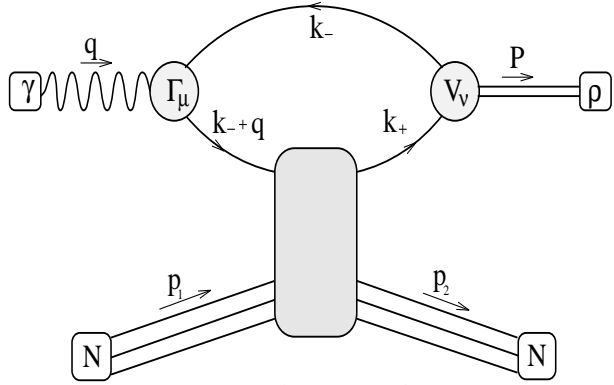,height=5.0cm,width=8.0cm}}}
\caption{
Impulse approximation to $\rho$-meson electroproduction current matrix 
element, \Eq{Current} with $t_{\mu \alpha \nu}(q,P)$ from \Eq{tman}.
}
\label{Fig:GRhoNN}
\end{figure}

Application of the quark-nucleon Pomeron-exchange interaction, defined
in \Eq{PomDef}, to $\rho$-meson electroproduction leads to the 
diagram illustrated in \Fig{Fig:GRhoNN}. The corresponding current
matrix element is
\ba
\lefteqn{
\langle P \lambda_{\rho} ; p_2 m^{\prime} | J_{\mu}(q) 
| p_1 m \rangle = 
 2 t_{\mu \alpha \nu}(q,P) \,
\varepsilon_{\nu}^{\lambda_{\rho}}(P)  }
\nn\ \\ 
& & \times  [G(w^2,t) \; 3 \beta_u F_1(t)]
\, [\bar{u}_{m^\prime}(p_2) \gamma_{\alpha} u_m(p_1)] 
.
\label{Current}
\ea
Here: $u_m(p_1)$ and $\bar{u}_{m^\prime}(p_2)$ are, respectively, the
spinors for the incoming and outgoing nucleon; 
$t = - (p_1 - p_2)^2 \leq 0$;  
$\varepsilon_{\nu}^{\lambda_{\rho}}(P)$ is the polarization vector of
the $\rho$ meson; and the factor of 2 arises from the equivalence
of the two contributing diagrams under charge conjugation because the
quark-Pomeron-exchange vertex is even under charge conjugation.
The Pomeron-exchange parameterization $G(w^2,t) \; 3 \beta_u F_1(t)$
has been defined in Eqs.~(\ref{GDef}) and (\ref{PomDef}).  
The energy of the quark-nucleon Pomeron-exchange interaction is taken
to be  $w^2= -(q + \frac{1}{2}P + p_1)^2$,
which follows from the observation that the quark-loop integration is
sharply peaked about a momentum for which the vector-meson BS amplitude
takes its maximum value.
This approximation ensures that 
$t_{\mu \alpha \nu}(q,P)$ is a function of $q$ and $P$ only, 
thereby allowing one of the four integrations necessary in calculating 
$t_{\mu \alpha \nu}(q,P)$ to be carried out analytically. 
We have investigated the ramifications of this approximation and find
that the $W$, $t$ and $q^2$ dependences of the observables considered
herein are insensitive to it and that the magnitudes of
cross section are insensitive to within 10\%.

The amplitude $t_{\mu \alpha \nu}(q,P)$ in \Eq{Current},
describes the coupling of the photon and $\rho$ meson to the nucleon
via Pomeron exchange. 
It is analogous to $\Lambda_{\mu}(q,P)$ in \Eq{TPiN} 
for meson-nucleon elastic scattering. 
In impulse approximation, $t_{\mu \alpha \nu}(q,P)$ is given 
by 
\ba
\lefteqn{
t_{\mu \alpha \nu}(q,P) = \beta_u N_c e_0
{\rm tr} \intd{k}
S_u(k_{-}) \Gamma_{\mu}(k_{-},k_{-}+q) 
}
\nn\ \\
& & \times  S_u(k_{-}+q)  \g_{\alpha}  S_u(k_{+})   
V_{\nu}(k_{+},k_{-})
,
\label{tman}
\ea
where 
$k_{\alpha } = k + \sfrac{\alpha}{2} P$, 
$N_c = 3$, $e_0 = \sqrt{4 \pi \alpha_{\rm EM}}$ and $\beta_u$ is given
in \Tab{Tab:PomParam}. 
All of the elements in \Eq{tman} have been determined in previous
sections, hence, there are {\em no} free parameters in our calculation
of vector-meson  electroproduction. 
We now present the results of our model and compare them
to the data.

\begin{figure}[t]
\centering{ 
\mbox{
\epsfig{figure=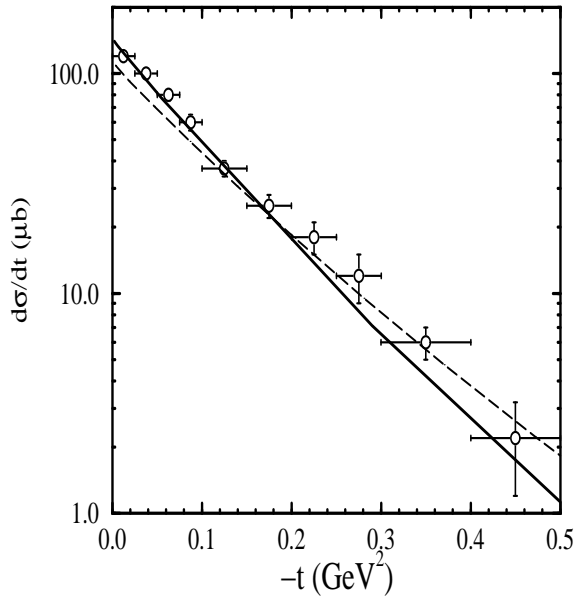,height=9.0cm,width=8.0cm}}}
\caption{
Our results for the differential cross section for $\rho$-meson
photoproduction ($q^2 = 0$) at $W = 80$~GeV.
The solid curve is the prediction of the present study and the dashed
curve is the result from \Ref{Pichowsky}. 
The data are from Ref.~\protect\cite{ZeusRhoPhoto}.
}
\label{Fig:dsigtZeus}
\end{figure}

We first consider the differential cross section for
$\rho$-meson photoproduction  at $W =$ 80~GeV. Our prediction is the
solid curve shown in \Fig{Fig:dsigtZeus}. 
Both the magnitude and $t$ dependence are in excellent agreement with
the data. In \Fig{Fig:dsigtZeus}, we also show the result (dashed
curve) from \Ref{Pichowsky}, obtained using 
the model parameter $\alpha_1 = $ 0.25~GeV$^{-2}$, 
which was taken directly from \Ref{DL87} and therefore, is 
{\em not} a self consistent determination within our model.
In the present work, the value $\alpha_1=$ 0.33~GeV$^{-2}$
is determined by the $t$ dependence of $\pi N$ elastic scattering 
as discussed in \Sec{Sec:ElasticPi}.   
Upon comparing the solid and dashed curves in \Fig{Fig:dsigtZeus}, we
observe that our new calculated value of 
$\alpha_1 = 0.33$ GeV$^{-2}$ has improved the agreement with data,
especially at small $t$.
This illustrates the importance of using a value of 
$\alpha_1$ that is consistent
with our confined-quark propagators and BS amplitudes.

\begin{figure}[t]
\centering{\ 
\mbox{\epsfig{figure=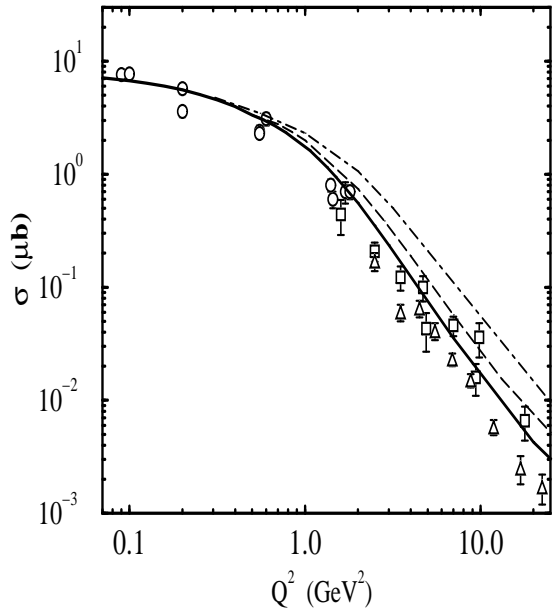,height=9.0cm,width=8.0cm}}}
\caption{
The $q^2$ dependence of the neutral $\rho$-meson electroproduction
cross section at $W =$ 15~GeV. The solid curve is the result of our
model. The data are from Ref.~\protect\cite{CHIO} (circles),
Ref.~\protect\cite{EMC} (squares) and Ref.~\protect\cite{NMC}
(triangles). 
The dashed and dot-dashed curves are the cross sections obtained from
our model, but with current-quark masses of  
$10$ and $25$ times greater than that given for $m_u$ in
\Tab{Tab:QuarkParam}.
}
\label{Fig:TotalRho}
\end{figure}
We now consider the $q^2$ dependence of the $\rho$-meson
electroproduction cross section. This is the most important result
obtained in this work. 
Our model elucidates the relation between the $q^2$ dependence of the  
electroproduction cross section and the quark substructure of the
vector meson. 
At present, the origin of the $q^2$ dependence of diffractive
electroproduction is still under debate.  We show here 
that, in our model, it arises {\em entirely} from the quark
substructure of the vector meson.

In \Fig{Fig:TotalRho}, the predicted $q^2$ dependence of the total
cross section is compared with the data. 
Our prediction for $W = 15$ GeV and $\epsilon = 0.5$ is shown as the
solid curve in \Fig{Fig:TotalRho}.  The agreement with the data is
excellent.   
We emphasize that {\em no} parameters were varied to achieve this
result. Such excellent agreement is unique to this model.  

Before we consider other electroproduction observables, 
let us illustrate some of the dynamics that gives rise to the
predicted $q^2$ dependence of the $\rho$-meson electroproduction cross
section (solid curve) shown in \Fig{Fig:TotalRho}.  
First, we observe that our prediction exhibits different behaviors in
the low-$q^2$ and high-$q^2$ region.   
For $q^2 <$ 1~GeV$^2$, the cross section falls off slowly with $q^2$, 
while for $q^2 >$ 1~GeV$^2$, it falls off more rapidly
and can be parameterized by the following form: 
\be
\sigma(W\!=\!15\;{\rm GeV},q^2 > 1 \;{\rm GeV}^2 ) = \sigma_0 
\left(\frac{q^2}{1\;\;{\rm GeV}^2}\right)^{a/2}
\label{q4falloff}
,
\ee
with $\sigma_0 = 2.047 \; \mu b$, and  $a = -4.070$, so that 
$\sigma(q^2 > 1\;{\rm GeV}^2) \propto 1/q^4$.
This value of $a$ is in good agreement with those obtained by
experiment: 
$a = -3.96 \pm 0.36$ from Ref.~\cite{EMC},
$a = -4.10 \pm 0.18$ from Ref.~\cite{NMC} and 
$a = -4.2 \pm 0.8$ from Ref.~\cite{ZeusRhoElectro}.
Theoretical predictions of $a$ in models that employ perturbative
methods typically find values in the range $a = -6$ \cite{Brodsky} 
and $a \approx - 4.8$ \cite{Martin}.

The dynamical origin of the $1/q^4$ dependence in our model
can be understood from the following considerations.
The domain of the integration momentum sampled in \Eq{tman} is
principally determined by the BS amplitude of the $\rho$ meson.  
This is because the BS amplitude provides most of the damping to
the integrand.
Therefore, the integrand is sharply peaked at
$k^2 \approx 0$, where the $\rho$-meson BS amplitude has its maximum
value. 
The integration can be approximated by evaluating the integrand at
$k^2 = 0$, in which case we find that the photon momentum, $q$, is
primarily concentrated in {\em one} of the quark propagators.  
(This was also observed in \Ref{DL87}.)
This quark propagator results in a $1/q^2$ dependence in \Eq{tman} for 
large $q^2$.  Since the cross section is proportional  to the square
of the hadron current (\ref{Current}), we have,
\be
 \sigma(W,q^2\gg q_0^2) \propto \frac{1}{1 + [q^2 / (q_0^2)_{\rho} ]^2}
,
\ee
where $(q_0^2)_\rho$ is a scale that determines the onset of the
$1/q^4$ behavior. This argument is independent of the
flavor of the quark or the particular vector meson under
consideration.   
It arises solely from the quark loop in \Eq{tman} and the fact that
the vector meson is a bound state with a BS amplitude peaked
at zero quark-antiquark relative momentum. 

The value of $q^2$, at which the transition to the $1/q^4$ behavior
occurs, depends on the scale $(q_0^2)_{\rho}$ which, in turn, depends
on the current-quark mass, $m_f$.  
To illustrate this, we allow the current-quark mass in the quark
propagator, $S_u(k)$, to take on values 10 and 25 times larger than
the value of $m_u =$ 5.1~MeV, given in \Tab{Tab:QuarkParam}.  
To remain self-consistent, we recalculate the parameters $a_{\rho}$
and $b_{\rho}$ of the $\rho$-meson BS amplitude in \Eq{VectorBSAmp} 
in order to maintain the values of the decay constants $f_{\rho}$ and 
$g_{\rho \pi \pi}$ in \Tab{Tab:VectorParam}.
The dashed (dot-dashed) curve shown in \Fig{Fig:TotalRho} is our
prediction of the $\rho$-meson electroproduction cross section with a
current $u$-quark mass 10 (25) times larger than $m_u =$ 5.1~MeV.  

We observe that all three curves in \Fig{Fig:TotalRho} converge to the 
same value at $q^2 = 0$.
This is due to the fact that our dressed quark-photon vertex,
$\Gamma^{\rm BC}_{\mu}(k,k^\prime)$, satisfies the WTI and the
Ward identity.  The Ward identity places very tight constraints on the
behavior of the cross section at $q^2=0$. 
This feature is also observed in studies of $\pi$-meson observables
where the unit normalization of the EM form factor arises from the
close connection between the normalization of the pion BS amplitude
and the Ward identity \cite{RobertsPi}.    

We also observe that for large values of $q^2$, all three curves in
\Fig{Fig:TotalRho} exhibit the {\em same} $1/q^{4}$ fall off. 
This $1/q^{4}$ behavior at high $q^2$ is a general feature of the
model, {\em independent} of the value of $m_f$.
However, the transition from a cross section that slowly decreases
with $q^2$ to one that falls off like $1/q^{4}$, 
occurs at a value of $(q_0^2)_\rho$ which {\em increases} with the
current-quark mass, $m_f$. 
This suggests that the electroproduction cross sections for
heavy-quark
vector mesons reach this $1/q^{4}$ fall off at larger values of $q^2$
than for those of light-quark vector mesons.   
This feature is illustrated in Secs.~\ref{Sec:VectorPhi} and
\ref{Sec:VectorJPsi} when we consider $\phi$- and $J/\psi$-meson
electroproduction. 

We emphasize that this dependence on the quark mass is a result of
having explicitly carried out the quark-loop integration of \Eq{tman}.
In studies that do not explicitly carry out the quark-loop
integration, it is necessary to introduce a quark-Pomeron-exchange
form factor in order to account for the $q^2$ dependence of the vector
meson electroproduction cross section \cite{DL87,Laget}.  
The introduction of this form factor destroys the relationship between
the scale of the transition, $(q^2_0)_\rho$, and the current-quark
mass, $m_f$. Furthermore, the value of $(q^2_0)_\rho$ depends on the
quark flavor (since it depends on $m_f$) which means that the
quark-Pomeron-exchange form factor is also flavor {\em dependent}
\cite{Laget}.   
Such flavor dependence violates our first assumption in
\Sec{Sec:Model} and would introduce additional, undetermined
parameters in the quark-nucleon Pomeron-exchange interaction.
The only flavor dependence in our approach (apart from the constant 
$\beta_f$) arises naturally from the dynamics of the intrinsic quark 
substructure of the vector meson. 

\begin{figure}[t]
\centering{\ 
\mbox
{\epsfig{figure=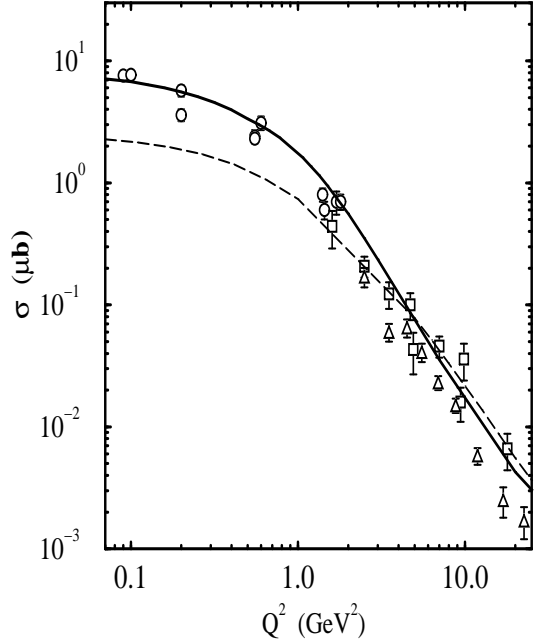,height=8.50cm,width=8.20cm}}}
\caption{
The effect of dressing the quark-photon vertex on the 
$\rho$-meson electroproduction cross section.
The solid curve is the result of our model with a dressed quark-photon
vertex.  
The dashed curve is the result of our model with a {\em bare}
quark-photon vertex. 
The data are the same as in \Fig{Fig:TotalRho}.
}
\label{Fig:TotalBC}
\end{figure}
Another important element in the determination of the $q^2$ dependence of
electroproduction is the dressed quark-photon 
vertex defined by Eq.~(\ref{GammaBC}).  We have already indicated that
the value of the diffractive electroproduction cross section, at $q^2 = 0$,
is tightly constrained by the fact that the quark-photon vertex employed
satisfies the Ward identity.   
As an estimate of its influence on the electroproduction
cross section, we replace the quark-photon vertex, 
$\Gamma_{\mu}(k,k^\prime)$, in \Eq{tman} with the bare vertex, 
$\gamma_{\mu}$.   
The resulting cross section is shown as a dashed curve in
\Fig{Fig:TotalBC}. 
It can be compared to the full calculation, shown as a solid curve. 
Clearly, the nonperturbative dressing of the quark-photon vertex is
essential in obtaining agreement with data for low $q^2$, where it
contributes significantly to the cross section. 

We would like to emphasize that although the effect of 
nonperturbative contributions to the quark-photon vertex is lessened in the
limit $q^2 \rightarrow \infty$, it does {\em not} vanish.  
It is a misapprehension to expect that at large $q^2$, the
perturbative limit, $\Gamma_{\mu}(k,k^\prime) \rightarrow
\gamma_{\mu}$, is approached. 
The quark-photon vertex, $\Gamma_{\mu}(k,k^\prime)$, only approaches
this perturbative limit when {\em both} quark momenta become large and
spacelike.  
From the above argument, we know that the quark-loop integration in  
\Eq{tman} receives its principal support when $k^2 \approx 0$ and
hence one quark propagator carries a large, spacelike momentum, 
$q \- \sfrac{1}{2} P$ (as illustrated by setting $k=0$ in
\Fig{Fig:GRhoNN}). 
However, the other quark propagator, which is attached to the
quark-photon vertex, carries momentum $\sfrac{1}{2}P$: 
this quark propagator is soft. 
Consequently, there is {\em no} value of $q^2$ for which the
quark-photon vertex can be replaced by $\gamma_{\mu}$.  
In the exclusive process of diffractive vector-meson
electroproduction, nonperturbative dressing of the quark-photon vertex
is always present, at {\em all} $q^2$.  

\begin{figure}[t]
\centering{\ 
\mbox
{\epsfig{figure=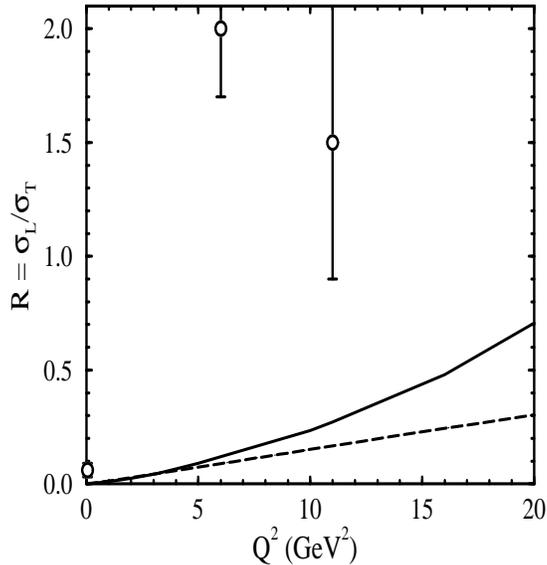,height=8.50cm,width=8.0cm}}}
\caption{
The $q^2$ evolution of the ratio of longitudinal to transverse cross
sections, $R(q^2)$, for $\rho$-meson electroproduction. 
The solid curve is the prediction of our model. 
The dashed line is the prediction of our model with a bare
quark-photon vertex, $\gamma_{\mu}$.  
The data points are estimated in Ref.~\protect\cite{NMC} and 
Ref.~\protect\cite{Warsaw} assuming $s$-channel helicity conservation. 
}
\label{Fig:SigmaLT}
\end{figure}
It is interesting to note that some observables are particularly
sensitive to the nonperturbative dressing of the quark-photon vertex, 
$\Gamma_{\mu}(k,k^\prime)$.  
As an example of this, in \Fig{Fig:SigmaLT}, we show the predicted
ratio of the longitudinal and transverse cross sections defined by 
\be
R(q^2) \equiv \frac{ \sigma_L }{\sigma_T}
.
\ee

This ratio has been examined by previous authors, using both nonperturbative
\cite{DL87} and perturbative models \cite{Dosch}. 
Although, their treatment of quark dynamics differ significantly from each
other,  and from that employed here, they yield  
results that are consistent with the data shown in \Fig{Fig:SigmaLT}.

The result obtained from our model is shown as the solid curve in
\Fig{Fig:SigmaLT}. 
The dashed curve is obtained from our model by replacing the dressed
vertex  with the bare quark-photon vertex, $\gamma_{\mu}$. 
We observe that nonperturbative dressing on the quark-photon vertex has a 
significant effect on $R(q^2)$ for all values of $q^2$.  

A comparison of $R(q^2)$ obtained in our model (solid curve) with the
data in \Fig{Fig:SigmaLT}, reveals a significant discrepancy.  If the
discrepancy is real, it might suggest that $R(q^2)$ is sensitive to
{\em transverse} contributions to the dressing of the quark-photon
vertex; i.e.\ to $\Gamma_{\mu}^{\rm T}(k,k^\prime)$. 
Such contributions have been neglected in this work, as discussed in
\Sec{Sec:ElasticPi}.   
By comparing the solid and dashed curves of
\Fig{Fig:SigmaLT}, we observe that $R(q^2)$ is very sensitive to the
nonperturbative dressing of the quark-photon vertex. 
So, although the inclusion of transverse contributions to the 
quark-photon vertex plays little role in most hadron observables,
$R(q^2)$ (a {\em ratio} of two cross sections, each approaching zero
with increasing $q^2$) may magnify such dynamical effects. 
This is an intriguing possibility;  the ratio $R(q^2)$ may provide a
new and important tool with which to investigate the dressing of 
the quark-photon vertex. Such an investigation is left for future
work. 

\subsection{Electroproduction of $\phi$ mesons}
\label{Sec:VectorPhi}
The $\phi$-meson electroproduction current matrix amplitude is
obtained from \Eq{Current} by relabeling $\rho \rightarrow \phi$.  
The associated amplitude, $t^{[\phi]}_{\mu \alpha \nu}(q,P)$, 
may be obtained from \Eq{tman} by 
relabeling $u \rightarrow s$ and multiplying by $\sqrt{\sfrac{2}{9}}$.
As in the case of $\rho$-meson electroproduction, all elements of the
calculation have been specified previously.
Therefore, there are no free parameters in the application of our
model to $\phi$-meson electroproduction.

\begin{figure}[t]
\centering{\ 
\mbox
{  
  \epsfig{figure=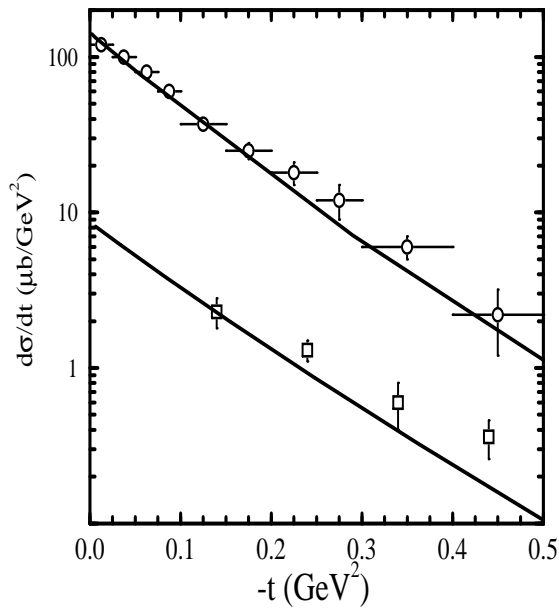,height=9.0cm,width=8.0cm}
}}
\caption{
The differential cross sections for 
photoproduction ($q^2 = 0$) of $\rho$ mesons (top
curve) and $\phi$ mesons (bottom curve) at $W=$ 80~GeV. 
The data are from Refs.~\protect\cite{ZeusRhoPhoto,ZeusPhiPhoto}.
}
\label{Fig:dsigtZeusPhi}
\end{figure}
In \Fig{Fig:dsigtZeusPhi}, we show the $\phi$-meson photoproduction 
differential cross section (lower curve) predicted by our model.  
It is in good agreement with the recent data.
In the same figure, the $\rho$-meson photoproduction cross section is
also displayed for comparison. 

\begin{figure}[t]
\centering{\
\mbox
{ 	
 \epsfig{figure=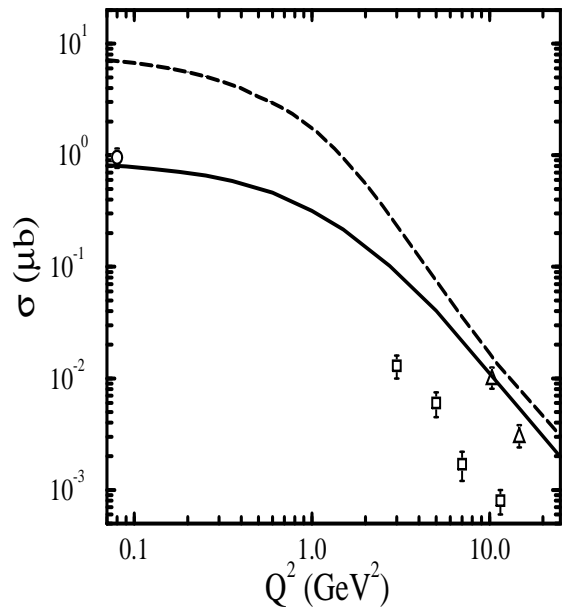,height=9.0cm,width=8.0cm}
}}
\caption{
The solid curve is our prediction for the $q^2$ dependence of the
$\phi$-meson 
electroproduction cross section at $W = $ 70~GeV. 
The dashed curve is our prediction for the $\rho$-meson
electroproduction cross section at $W =$ 15~GeV, shown here for
comparison. 
The $\phi$-meson electroproduction data are from 
\Ref{ZeusPhi} (triangles) and \Ref{NMC} (squares).
$\phi$-meson photoproduction ($q^2=0$) data from 
\Ref{ZeusPhiPhoto} (circles) 
are also shown at $q^2 \approx$ 0.08~GeV$^2$. 
}
\label{Fig:TotalPhi}
\end{figure}

The characteristics that distinguish between the quark substructure of
the $\phi$ and $\rho$ mesons can be illustrated by comparing the $q^2$
dependence of their electroproduction cross sections.  
The $q^2$ dependence of the predicted $\phi$-meson electroproduction
cross section for $W = 70$ GeV and $\epsilon = 0.5$ is given in
\Fig{Fig:TotalPhi} as a solid curve. 
We have also included our $\rho$-meson electroproduction results
(dashed curve) for comparison. 
We see that for large values of $q^2$, the $\rho$- and 
$\phi$-meson cross sections obey the asymptotic, power law $\approx
1/q^4$.  
In $\phi$-meson electroproduction, the transition to this asymptotic
region occurs at $(q_0^2)_{\phi} \approx 2$ GeV$^2$.  
The larger value of $(q^2_0)_{\phi}$ relative to that of
$(q^2_0)_{\rho} \approx$ 1~GeV$^2$ reflects the difference in
magnitude between the relevant scales of the $u$ and $s$-quark
propagators, namely $m_u$ and $m_s$. 
This result was anticipated from our discussion in
\Sec{Sec:VectorRho}. 
Furthermore, this result entails that for large enough $q^2$, the
ratio of the $\rho$- and $\phi$-meson electroproduction cross
sections, $\sigma^{[\rho]} / \sigma^{[\phi]}$ approaches a constant.
 
While our predictions presented in \Fig{Fig:TotalPhi} 
are in agreement with the recent data from HERA 
(circles and triangles),  
there is a considerable discrepancy between our results and the data 
reported by the NMC (squares) \cite{NMC}.
(The NMC data were measured at the lower energy $W \approx$ 14~GeV. 
However, the discrepancy between our prediction and the NMC data
persists even when the $W$ dependence of our model is accounted for.) 
The normalization of these data seem abnormally low when compared to
the recent data from HERA. 
This may be due to the fact that these $\phi$-meson
electroproduction data were obtained by averaging over several 
{\em different} nuclear targets (deuterium, carbon and calcium). 
If the nuclear effects are not properly accounted for, then a direct 
comparison of these data to our results on a nucleon target, is
meaningless.   
There is evidence from FNAL experiment E665, that the NMC
data for $\rho$-meson electroproduction may also be low by a factor of
2 \cite{Don}.  A careful comparison of our prediction for the
$\rho$-meson electroproduction cross section (solid curve in
\Fig{Fig:TotalRho}) and the NMC data for this process (triangles in
\Fig{Fig:TotalRho})  reveals that these data are a factor of 2 lower
than our prediction. We therefore view these data with caution.

It is impossible to reconcile the NMC data for $\phi$-meson
electroproduction with our model. 
Agreement with these data requires that the momentum scale
$(q^2_0)_{\phi}$, which determines the onset of the $1/q^4$ fall off
for the $\phi$-meson electroproduction cross section, would have to be
the same as $(q^2_0)_{\rho}$ for $\rho$-meson electroproduction. 
Based on our analysis of the dependence of $(q^2_0)_{\rho}$ on $m_f$ 
in \Sec{Sec:VectorRho}, the requirement that 
$(q^2_0)_{\phi} \approx (q^2_0)_{\rho}$
implies $m_s \approx m_u$. 
On the contrary, since $m_s \gg m_u$, the $\phi$-meson
electroproduction cross section should reach the asymptotic $1/ q^{4}$
region {\em later} than that of the $\rho$ meson. 
We conclude that the NMC data (triangles) in \Fig{Fig:TotalPhi} is not 
simply related to $\phi$-meson electroproduction on a {\em nucleon}
target, and hence such an interpretation of these data is erroneous.

If these NMC data are truly to be interpreted as a measurement of $\rho$- and
$\phi$-meson electroproduction on a {\em nucleon} then a comparison with
recent data from HERA (at higher energy but similar $q^2$) would entail that,
at large-$q^2$, electroproduction cross sections exhibit an energy dependence
much stronger than that observed in low-$q^2$ diffractive processes
\cite{Warsaw}.  However, our observation that the NMC data are consistently
low suggests such a conclusion to be incorrect.

\begin{figure}[t]
\centering{ 
\mbox
{  
\epsfig{figure=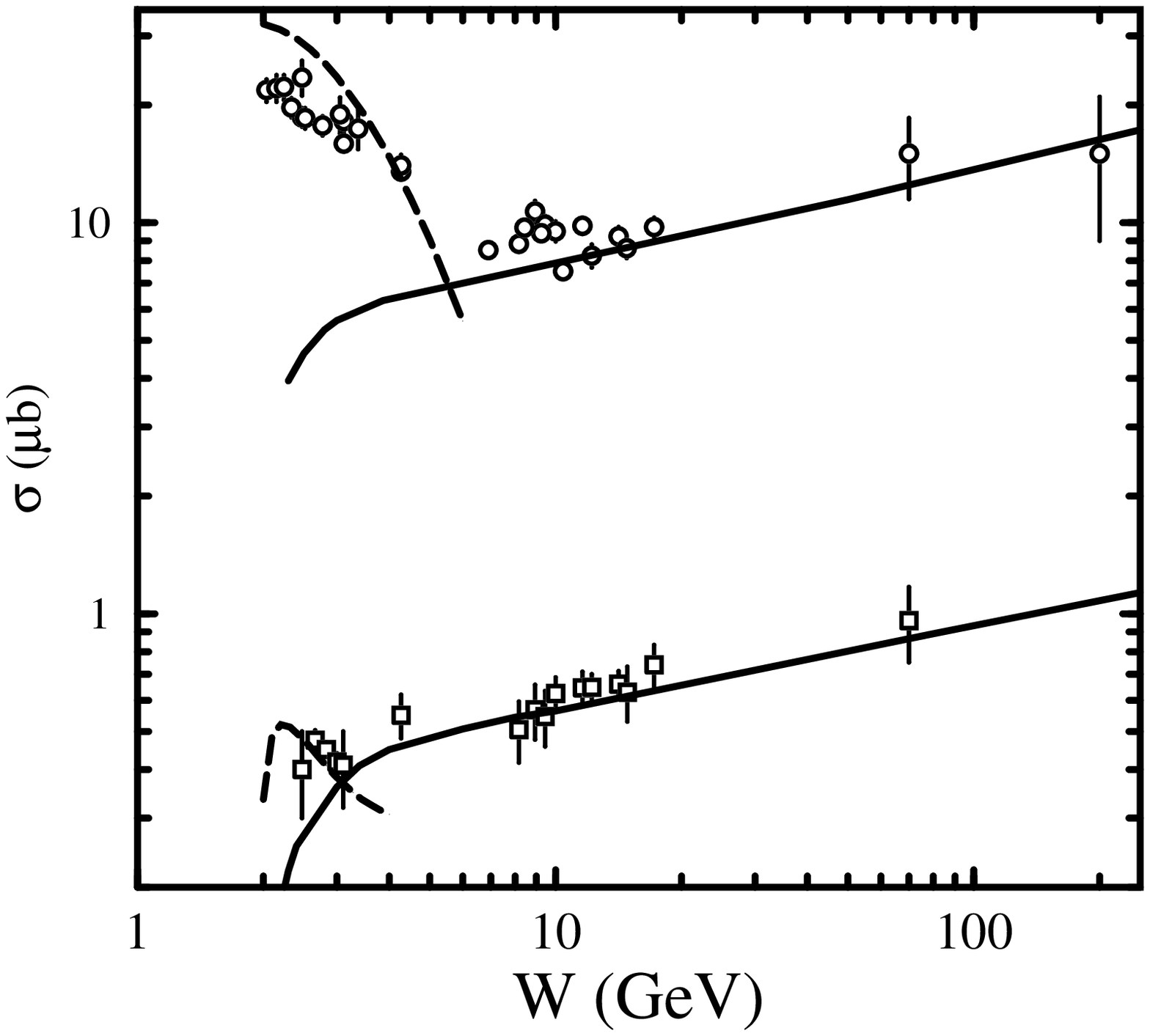,height=9.0cm,width=8.0cm}
}}
\caption{
Energy dependence of $\rho$- (top) and $\phi$-meson (bottom)
photoproduction cross sections. The solid curves are the predictions
from our quark-nucleon Pomeron-exchange interaction.  The dashed
curves are the predictions of the meson-exchange model discussed in
the text. 
The $\rho$-meson data (triangles) are from 
\Refs{ZeusRhoPhoto,CHIO,Ballam,Struczinski,Egloff,Aston}.
The $\phi$-meson data (squares) are from 
\Refs{ZeusPhiPhoto,Ballam,Egloff,Barber83}. 
}
\label{Fig:SigmaW}
\end{figure}

We now investigate the consequence of our quark-nucleon
Pomeron-exchange interaction in determining the energy dependence 
of vector meson electroproduction cross sections. 
The predicted energy dependence of $\rho$-meson and $\phi$-meson
photoproduction are shown as the upper and lower solid curves in
\Fig{Fig:SigmaW}, respectively. 
No parameters have been varied to obtain these results.  
Within our model, the rate of increase with energy is completely
determined by the value of $\alpha_0$.  This is due to the
assumption that the energy dependence of $G(s,t)$ in \Eq{GDef} can be 
calculated from the external legs, and hence the quark-loop
integration in \Eq{tman} only generates a $t$ dependence.
The results shown in \Fig{Fig:SigmaW} and previous figures indicate
that this assumption is valid and sufficient to describe light-quark
meson-hadron interactions.  
However, $J/\psi$-meson electroproduction has
a much steeper $W$ dependence which {\em cannot} be completely
accounted for with $\alpha_0 \approx 0.10$.  This is discussed further
in the next section. 

At low energies, large-$t$ contributions become important and
other (non-diffractive) exchange mechanisms must be included to
describe the $\rho$- and $\phi$-meson electroproduction data.  
To demonstrate this, we have carried out a simple calculation using
the lowest order meson-exchange contributions derived from the
phenomenological Lagrangian of Ref.\cite{SatoLee}. 
For $\phi$-meson photoproduction, we only consider $\pi$ and $\eta$
exchanges and use the parameters from the Bonn potential \cite{Bonn}
and experimental decay constants. 
We introduce a cut-off function $\Lambda^2/(\Lambda^2+ k^2)$ with
$\Lambda = $ 1~GeV, so that the magnitudes of the low-energy cross
sections agree roughly with experiment.  
The resulting cross sections are shown as dashed curves in
\Fig{Fig:SigmaW}. 

It is interesting to note that the meson-exchange contributions
decrease with energy so that at sufficiently high energies Pomeron
exchange dominates. 
We see from \Fig{Fig:SigmaW} that the $W$ dependence of meson exchange
is strikingly different than that of Pomeron exchange.  
This difference suggests that these two exchanges may be modeling very
different aspects of QCD. 
For example, one might view meson exchange as a phenomenological
representation of the exchange of correlated quark-antiquark pairs. 
Its strength, therefore, depends strongly on the flavor structure of
the hadrons involved. 
Therefore, it is not then surprising that meson-exchange contributions
to $\phi$-meson photoproduction are significantly less than those to
$\rho$-meson photoproduction; 
the lack of valence $s$-quarks in the nucleon tends to suppress direct
quark exchanges. 
The fact that Pomeron exchange contributes similarly to both of these
processes, independent of the flavor of the quark substructure of the
hadrons involved (as shown in \Fig{Fig:SigmaW}), is suggestive that
the underlying mechanism of Pomeron exchange is multiple-gluon
exchange.  
This idea was proposed long ago in Refs.\cite{Low,Nussinov} and is
supported by much experimental evidence.   
Our results in \Fig{Fig:SigmaW} provide additional evidence in support
of this notion. However, further theoretical work is necessary to make
compelling the identification between Pomeron exchange and
multiple-gluon exchange. 

\begin{figure}[t]
\centering{ 
\mbox
{
\epsfig{figure=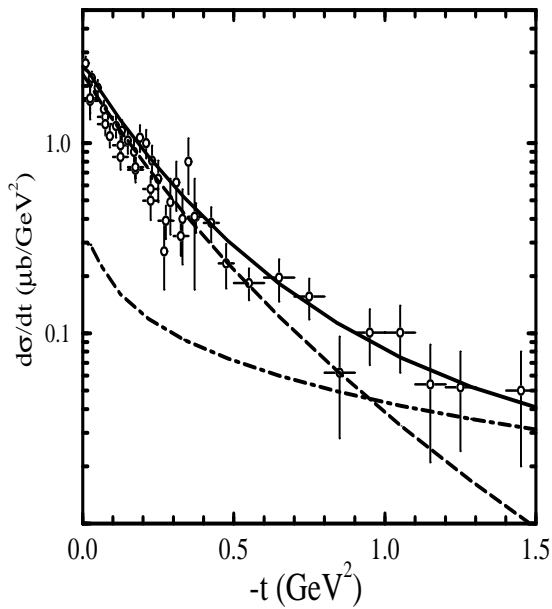,height=9.0cm,width=8.0cm}
}}
\caption{
The differential cross section for $\phi$-meson photoproduction 
($q^2 =0$) 
for  $3.0 \leq W \leq$ 3.5~GeV. 
The dashed curve is the contribution due to the
Pomeron-exchange interaction and the dot-dashed curve is the
contribution due to $\pi$ and $\eta$ exchange at $W=$ 3.0~GeV.
The solid curve is the sum of these.
The data are from Refs.~\protect\cite{Barber83,Barber78,Behrend}.
}
\label{Fig:dsigt3}
\end{figure}

As a second illustration of the differences between Pomeron exchange
and meson exchange in $\phi$-meson photoproduction, we present our
predictions for the differential cross section at $W =$ 3~GeV in
\Fig{Fig:dsigt3}. 
We observe that Pomeron exchange (dashed curve) is dominant in the
forward peak while $\pi$ and $\eta$ exchanges (dot-dashed curve)
contribute a  flatter background. 
This is in accord with our expectation that Pomeron exchange is 
responsible for the diffractive, forward peak observed in these cross 
sections. 

The absence of valence-$s$ quarks in the nucleon tends to suppress
meson-exchange contributions to $\phi$-meson electroproduction.
This provides the opportunity for experimental facilities, such as
TJNAF, 
to probe the interplay between Pomeron exchange and meson exchange in
processes like $\phi$-meson photoproduction.  
In particular, spin observables may be used to extract details
concerning the Dirac structure of the quark-Pomeron-exchange coupling.
Such studies could provide new insights into the QCD dynamics
underlying Pomeron exchange.

\subsection{Electroproduction of $J/\psi$ mesons}
\label{Sec:VectorJPsi}

The final process we consider is $J/\psi$-meson electroproduction. 
Although data is scarce and our treatment of the $c$-quark propagator
is necessarily simplistic, it is nonetheless, instructive to perform
an exploratory study of this process in our model.

It is observed that the $J/\psi$-meson photoproduction cross section
rises more sharply with energy than do $\rho$- and $\phi$-meson
photoproduction cross sections. 
The energy dependence of our model is determined by the parameter
$\alpha_0$, and is, by construction, the same for all diffractive
processes. 
Therefore, we anticipate the need to alter the value of $\alpha_0$ to 
properly account for this energy dependence.  
We also find that the $t$ dependence of $J/\psi$ photoproduction
predicted by our model is too steep. In this section, we allow a
minimal flavor dependence of the model parameters necessary to provide
a good description of $J/\psi$-meson electroproduction.  
Although the model does not predict the observed $J/\psi$
photoproduction cross sections, we argue that our model of the quark
substructure of the $J/\psi$ meson is sufficient and that it is our
assumption of a flavor-dependent Pomeron-exchange amplitude that must
be relaxed to obtain agreement with the data.

The $c$ quark has not been studied to the extent that the $u$, $d$ and
$s$ quarks have.  Therefore, we develop a simple model $c$-quark
propagator and 
consider the EM decay, $J/\psi \rightarrow e^+ e^-$, to determine the
$J/\psi$-meson BS amplitude.  

Numerical studies \cite{Jain} suggest to us that the $c$-quark
propagator can be more simply parameterized than the light-quark
propagators. 
The heavy mass of the $c$ quark, $m_c \approx$ 1.2~GeV, provides the
dominant scale in the propagator.  This observation supports the use
of a model $c$-quark propagator that exhibits confinement, reduces to
the correct asymptotic form, required by perturbative QCD, but has
only one momentum scale, $m_c$. We, therefore, employ the model
propagator: 
\be
S_c(k) = ( - i \g \cdot k + m_c ) 
\frac{1 - e^{-(1+k^2/m_c^2)}}{k^2 + m_c^2}
\label{cQuarkProp}
.
\ee

There are fewer phenomenological constraints on the BS amplitude of
the $J/\psi$ meson since there is no $J/\psi$ decay analogous to  
$\rho \rightarrow \pi \pi$; we have only $J/\psi \rightarrow e^+ e^-$
to fix the parameters in the BS amplitude.  
To reduce the number of parameters in the vector-meson BS amplitude
given in \Eq{VectorBSAmp}, we set $c_{J/\psi} = 0$, thereby
eliminating two of the parameters and the $1/k^2$ behavior of the
amplitude.  The BS amplitude is 
then,
\be
V_{\mu}(p,p^\prime) = \frac{1}{N_{J/\psi}} e^{-k^2 / a_{J/\psi}^2}
\left[\gamma_{\mu}+\frac{P_{\mu}\gamma\cdot P}{m_{J/\psi}^2} 
\right],
\label{VJPsi}
\ee
where $P = p \- p^\prime$ is the total momentum of the $J/\psi$ meson
and $k = \sfrac{1}{2}(p \+ p^\prime)$ is the relative momentum between
the quark and antiquark.
The sole parameter, $a_{J/\psi}$, is determined by reproducing the
experimental decay width of $J/\psi \rightarrow e^+e^-$, using the
procedure discussed in \Sec{Sec:Vector}.
The BS normalization, $N_{J/\psi}$, is obtained
from \Eq{VectorBSN}.  
The resulting value of $a_{J/\psi}$ and the calculated EM decay
constant are given in \Tab{Tab:VectorParam}.
The $c$-quark-Pomeron-exchange coupling constant, $\beta_c$,
cannot be fixed from hadron-hadron elastic scattering since there are
no data for the scattering of $c$-quark mesons from a nucleon target.
Therefore, we choose the value of $\beta_c$ such that the model
reproduces the magnitude of the $J/\psi$-meson photoproduction ($q^2 =
0$) cross section at $W=$ 100~GeV from \Ref{ZeusJPsiPhoto}.  
In contrast to $\rho$- and $\phi$-meson electroproduction,
the overall normalization of the $J/\psi$-meson electroproduction
cross section is {\em not} a prediction of the model.   

Upon substituting the BS amplitude of \Eq{VJPsi} and the $c$-quark
propagator, \Eq{cQuarkProp}, into \Eq{tman} and multiplying by
$\sqrt{\sfrac{8}{9}}$ (which arises from the flavor structure of the
$J/\psi$ meson), we obtain the photon-$J/\psi$-Pomeron-exchange
transition amplitude, $t^{[J/\psi]}_{\mu \alpha \nu}(q,P)$.  
This amplitude is then substituted into \Eq{Current} to calculate the 
cross section for $J/\psi$-meson electroproduction.

\begin{figure}[t]
\centering{\ 
\mbox
{\ \epsfig{figure=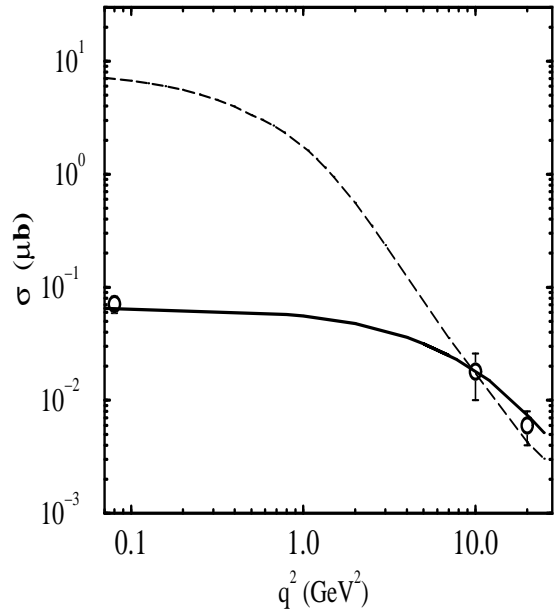,height=9.0cm,width=8.0cm}}}
\caption{
The total $J/\psi$-meson electroproduction cross section at $W =$
100~GeV (solid curve) 
and total $\rho$-meson electroproduction cross section at  
$W =15$~GeV (dashed curve).
The $J/\psi$-meson data are from
Refs.~\protect\cite{ZeusJPsiPhoto,H1RhoJPsi}. 
}
\label{Fig:TotalJPsi}
\end{figure}

The predicted $q^2$ dependence of the $J/\psi$-meson electroproduction
cross 
section is the solid curve shown in \Fig{Fig:TotalJPsi} at $W=$
100~GeV. 
Our prediction is in excellent agreement with the data.
It is important to emphasize that {\em no} parameters were adjusted to 
obtain this result.  As in the case of $\rho$- and $\phi$-meson
electroproduction, the $q^2$ dependence of the $J/\psi$
electroproduction 
cross section arises solely from the quark-loop integration and hence is a
{\em prediction} of our model. 

In the same figure, for comparison, we also display our results for
$\rho$-meson electroproduction.
The presence of heavy $c$ quarks dramatically changes the $q^2$
dependence of the cross section. 
(This was anticipated from our analysis of the $q^2$
dependence of $\rho$ and $\phi$-meson electroproduction in
Secs.~\ref{Sec:VectorRho} and \ref{Sec:VectorPhi}.)
Remarkably, although the $J/\psi$-meson photoproduction ($q^2= 0$)
cross section is 2 orders of magnitude lower than that for the $\rho$
meson, they are equal at $q^2 \approx$ 15~GeV$^2$.
(We note that the curves shown in \Fig{Fig:TotalJPsi}
are evaluated at {\em different} energies.  When the energy dependence
of the $\rho$-meson cross section is taken into account, the two
curves intersect at $q^2 \approx$ 15~GeV$^2$.) 
This surprising result arises naturally in our model as a consequence
of the dynamical treatment of the quark loop which was shown, in
\Sec{Sec:VectorRho}, to depend on the current-quark mass, $m_f$.  
Future experimental data for $J/\psi$-meson electroproduction,
especially at moderate $q^2$, would be helpful to confirm our
predictions. 

\begin{figure}[t]
\centering
{\ \epsfig{figure=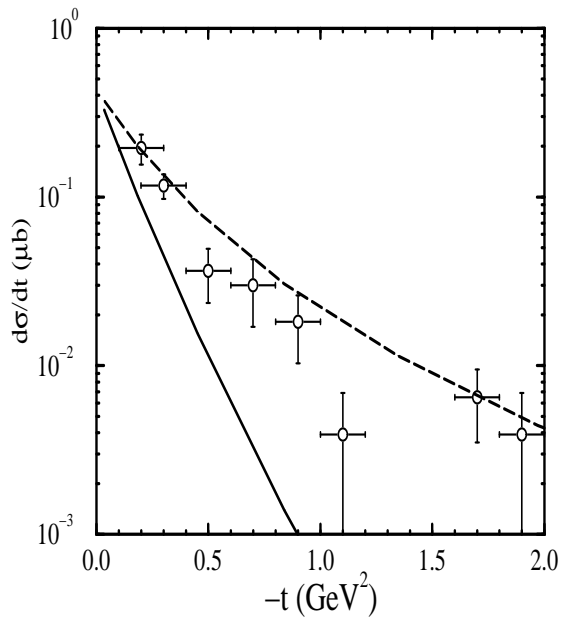,height=9.0cm,width=8.0cm}}
\caption{
The differential cross section for $J/\psi$-meson photoproduction
($q^2=0$). The solid curve is our result at $W =$ 40~GeV. 
The dashed curve is an estimation of the contribution to the
differential cross section from the quark substructure of the $J/\psi$
meson and nucleon. The data are from \Ref{ZeusJPsiPhoto}.
}
\label{Fig:dsigtJPsi}
\end{figure}

In \Fig{Fig:dsigtJPsi}, our prediction of the differential cross
section for $J/\psi$-meson photoproduction is compared to the data of 
\Ref{ZeusJPsiPhoto}.
The resulting solid curve fails to reproduce the observed $t$
dependence of the data.
In our model, there are three sources of $t$ dependence: the nucleon
form factor, $F_1(t)$ in \Eq{PomDef}, the Pomeron-exchange amplitude,
$G(s,t)$ in \Eq{GDef}, and the photon-$J/\psi$-Pomeron-exchange
coupling, $t^{[J/\psi]}_{\mu \alpha \nu}(q,P)$. 

In this exclusive process, the requirement that a $J/\psi$ meson is
produced and the nucleon does {\em not} break up, introduces a $t$
dependence.   
This $t$ dependence should be present in any model.
We can estimate the effect of these two constraints on the $t$
dependence by setting $\alpha_1 = 0$ in \Eq{GDef}.  The resulting
cross section is shown as a dashed curve in \Fig{Fig:dsigtJPsi}.
This curve can be represented as the net $t$ dependence arising from
the quark substructure of the $J/\psi$ meson 
($t^{[J/\psi]}_{\mu \alpha \nu}(q,P)$ from \Eq{tman})
and the nucleon ($\approx F_1(t)$ in \Eq{PomDef}).   
As such, it represents a bound on the $t$ dependence of the cross
section. 
The data in \Fig{Fig:dsigtJPsi} are consistent with this
interpretation in that they lie below the predicted bound.

Furthermore, in \Sec{Sec:VectorRho}, we have shown that the $q^2$
dependence of vector-meson electroproduction arises entirely from the
quark substructure of the vector meson.  
The fact that our model readily reproduces the observed $q^2$
dependence $J/\psi$ electroproduction (see \Fig{Fig:TotalJPsi})
suggests that the amplitude $t^{[J/\psi]}_{\mu \alpha \nu}(q,P)$ 
provides an
adequate description of the quark substructure of the $J/\psi$ meson. 
Therefore, we believe that the discrepancy between our result (solid
curve) and the data in \Fig{Fig:dsigtJPsi} is due to having employed
the value $\alpha_1 = $ 0.33~GeV$^{-2}$ obtained from our study of
$\pi N$ elastic scattering.  
We suspect that the assumed flavor-independence of the
parameters, $\alpha_0$ and $\alpha_1$, is sufficient to describe
diffractive processes involving light quarks, but that it is too
restrictive to provide agreement with the data for $J/\psi$-meson
electroproduction. 
By relaxing this constraint, the data in \Fig{Fig:dsigtJPsi}, for $t
\leq -1.3$ GeV$^2$, are reproduced with $\alpha^c_1 = 0.10$
GeV$^{-2}$. 

\begin{figure}[t]
\centering
{\ 
\epsfig{figure=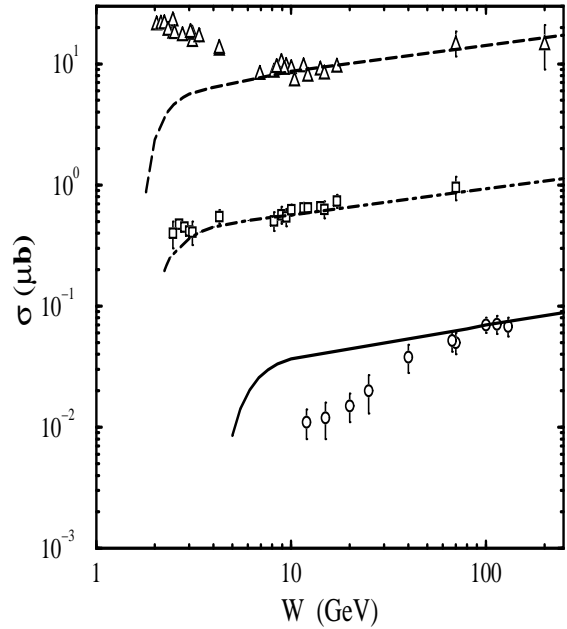,height=9.0cm,width=8.0cm}
}
\caption{
The predicted energy evolution of vector-meson photoproduction 
($q^2 = 0$) for $\rho$ (dashed curve), 
$\phi$ (dot-dashed curve) and $J/\psi$ mesons (solid curve).  
The data are from 
Refs.~\protect\cite{Ballam,ZeusJPsiPhoto,H1RhoJPsi}.
}
\label{Fig:SigmaWJPsi}
\end{figure}

As indicated earlier, the energy dependence of $J/\psi$
photoproduction is much stronger than the other processes we have
considered. 
Our results for $J/\psi$-meson photoproduction are plotted versus $W$
in \Fig{Fig:SigmaWJPsi} along with our results for the $\rho$ and
$\phi$ mesons.  
The $W$ dependence obtained from our model for $J/\psi$
photoproduction is not in agreement with the data. 
However, by relaxing the flavor-independence assumption in the model,  
one can account for this steep energy dependence.  The data for
$J/\psi$-meson photoproduction shown in \Fig{Fig:SigmaWJPsi} (circles)
are reproduced by choosing $\alpha_0^c = 0.42$.  

It has been suggested \cite{Jenkovszky} that the discrepancy between the
$W$ and $t$ dependence of $J/\psi$ electroproduction and that of $\rho$-
or $\phi$-meson electroproduction could be explained if the energies at
which current experiments are conducted are not yet in the asymptotic
region that would be well-described by Pomeron exchange.
Further experimental data on the diffractive electroproduction of the 
$J/\psi$ meson at higher energies would help to test this 
hypothesis. 

This concludes our discussion of $J/\psi$-meson electroproduction. 
We find that we can obtain good agreement with all data provided only
that we relax our model assumption that the Pomeron-exchange
parameters be independent of the flavor of the quarks involved.

\section{Summary and conclusions}
\label{Sec:Conclusion}

In this paper, we have developed a model quark-nucleon
Pomeron-exchange interaction and employed it in a study of the role of
the quark substructure of mesons in meson-nucleon elastic scattering
and exclusive vector-meson electroproduction on nucleons.  
The quark substructure of the mesons is described in terms of 
relativistic quark-antiquark Bethe-Salpeter amplitudes within a
framework developed from phenomenological studies of the
Dyson-Schwinger equations of QCD. 
The model developed herein provides a framework in which some of the
crucial aspects of nonperturbative-quark dynamics can be explored in
diffractive processes.  
As such, this work represents a first attempt to separate the effects
of hadron substructure from the effects of Pomeron exchange on the
observables of diffractive processes. 
This separation is a necessary step toward the development of an 
understanding of the quark and gluon dynamics underlying Pomeron
exchange. 

The quark-nucleon Pomeron-exchange interaction, introduced in
\Sec{Sec:Model}, is defined in terms of 4 parameters.   
They are determined by requiring that the model reproduce 
$\pi N$ and $K N$ elastic scattering data. 
The model interaction is then used to predict the electroproduction of
$\rho$ and $\phi$ mesons on nucleons.  
We find that the model successfully describes both the slopes and
magnitudes of the differential cross sections. 
The predicted energy and $q^2$ dependence of the $\rho$- and
$\phi$-meson photoproduction cross sections are in excellent agreement
with the data.    

For these process, Pomeron exchange provides the 
dominant contribution for large energies, but as the energy is
decreased, other mechanisms come into play.  
Meson exchanges contribute significantly at energies $W \leq $
6~GeV for $\rho$-meson electroproduction and $W \leq$ 3~GeV for
$\phi$-meson electroproduction. 
If meson exchange can be identified with correlated
quark-antiquark exchange then the dramatically different behaviors
predicted from the meson- and Pomeron-exchange mechanisms support the
notion that the dynamics underlying Pomeron exchange would be best
described in terms of gluon degrees of freedom. 
Future work will address this intriguing possibility.

The observed asymptotic $1/q^4$ dependence of vector-meson
electroproduction cross sections arises naturally from the quark
substructure of the vector meson. 
Our predictions are different from those of other authors
and are in agreement with the recent data from
HERA for $\rho$-, $\phi$- and $J/\psi$-meson diffractive
electroproduction. 
We show that the scale that determines the onset of this asymptotic
behavior is determined by the current-quark mass of the quarks inside
the vector meson. 
This work is the first to elucidate the relationship between the
$q^2$ dependence of diffractive electroproduction cross sections and
the current-quark mass. 
This important observation explains the dramatic differences between
the $q^2$ dependence seen in the electroproduction cross sections
of the various vector mesons.

We also find that the normalization of both $\rho$- and $\phi$-meson
electroproduction data from NMC are too low to be accounted for by our
model.  
This is based on our analysis, in \Sec{Sec:Vector}, 
of the role played by the current-quark mass in determining the onset of the
asymptotic $1/q^4$ behavior in vector-meson electroproduction. 
This finding is particularly important because these data have been used to
suggest that large-$q^2$ electroproduction cross sections have a stronger
energy dependence than that which is observed in low-$q^2$
electroproduction and elastic hadron-hadron scattering.
However, our results suggest that these NMC data do not represent an
accurate measure of vector meson electroproduction on a {\em nucleon}, 
hence such a conclusion can not be drawn. 

In an exploratory investigation of $J/\psi$-meson electroproduction,
we find that although the $q^2$ dependence of the cross section is
correctly predicted by our model, the observed energy dependence is
much steeper and the $t$ dependence much flatter than those we
obtain. 
This discrepancy between our model and the experimental data can be
removed by allowing a flavor dependence in the quark-nucleon
Pomeron-exchange interaction. 

In this work, we have given an indication of the important role played
by the quark substructure of hadrons in exclusive, diffractive
processes. 
In particular, the $q^2$ dependence of electroproduction cross
sections is a sensitive probe of the dynamical evolution of the
confined quark from the domain in which nonperturbative dressing
effects are important to that in which it behaves as a current quark. 
We also find that the minimal dressing of the quark-photon vertex,
required by the Ward-Takahashi identity, keeps the exclusive
vector-meson electroproduction cross section from reaching a
na\"{\i}ve perturbative-QCD limit as $q^2$ becomes large.
Hence, there are {\em always} some nonperturbative contributions to
these exclusive processes.
One observable that is particularly sensitive to the nonperturbative
dressing of the quark-photon vertex is the ratio of longitudinal to
transverse electroproduction cross sections, $R$.  
Future experimental measurements of $R$ will provide an
important tool with which to investigate the importance of the
nonperturbative dressing of the quark-photon vertex.

In conclusion, we have developed a model quark-nucleon
Pomeron-exchange interaction that successfully describes the extensive
data for diffractive processes on the nucleon.
A framework developed from studies of the Dyson-Schwinger equations of
QCD was employed to describe the quark substructure of mesons. 
Our results indicate that such a framework can provide a uniformly
good description of exclusive processes at both low and high energies. 
We argue that the quark substructure of vector mesons plays a central
role in determining the behavior of the electroproduction cross
sections considered, particularly the $q^2$ dependence of the cross
section. 
Our results are consistent with the notion that Pomeron exchange may
be identified with a multiple-gluon exchange mechanism within QCD.   
However, further work is required to make this possible identification
compelling.  
Such an investigation requires a careful treatment of the
quark 
substructure of hadrons.     


{\bf Acknowledgments \hspace*{0.50cm}}
The authors are grateful to C.D.~Roberts for many stimulating
conversations and helpful suggestions.
This work is supported by the U.S. Department of
Energy, Nuclear Physics Division, under contract W-31-109-ENG-38.
M.A.P. also received support from the Division of Educational Programs
at Argonne National Laboratory and the Dean of Graduate Studies of the
Faculty of Arts and Sciences of the University of Pittsburgh. 
The computations described herein were carried out using resources 
at the National Energy Research Scientific Computing Center.


\end{document}